# Uncertainty of current understanding regarding OBT formation in plants

A. Melintescu[*], D. Galeriu

"Horia Hulubei" National Institute for Physics and Nuclear Engineering, Department of Life and Environmental Physics, 30 Reactorului St., P.O. Box MG-6, Bucharest – Magurele, Romania, RO-077125

[*] Corresponding author: Phone: +40 21 4042359; Fax: +40 21 4574440; email: ancameli@ifin.nipne.ro, melianca@yahoo.com

**Abstract**

Radiological impact models are important tools that support nuclear safety. For tritium, a special radionuclide that readily enters the life cycle, the processes involved in its transport into the environment are complex and inadequately understood. For example, tritiated water (HTO) enters plants by leaf and root uptake and is converted to organically bound tritium (OBT) in exchangeable and non-exchangeable forms; however, the observed OBT/HTO ratios in crops exhibit large variability and contradict the current models for routine releases. Non-routine or spike releases of tritium further complicate the prediction of OBT formation. The experimental data for a short and intense atmospheric contamination of wheat are presented together with various models' predictions. The experimental data on wheat demonstrate that the OBT formation is a long process, it is dependent on receptor location and stack dynamics, there are differences between night and day releases, and the HTO dynamics in leaf and ear is a very important contributor to OBT formation.

## 1. Introduction

Tritium ($^3$H) is present in the environment as a result of both natural and anthropogenic sources. Large quantities of tritium are currently produced in heavy water reactors and fuel reprocessing plants, and it is anticipated that the development of fusion energy will increase environmental releases. Romania develops nuclear energy using Canadian heavy water reactors and two CANDU 6 units are in operation having large tritium loads. The past and present Romanian research studies cover environmental tritium monitoring (Paunescu et al., 2012), environmental tritium modelling (Barry et al., 1999), and it was pointed out that tritium has to be considered as a special radionuclide (Galeriu and Melintescu, 2010). Tritiated water (HTO) and tritiated gas (HT) are the major forms initially released in environment and the organic forms (tritiated methane or tritiated formaldehyde) have minor releases (Amano, 1995; Kakiuchi et al.2002).

In the process of photosynthesis, plants produce organic matter using solar light as energy source, from carbon dioxide from the air, nutrients from soil, and water from soil or air. Because tritium is larger than hydrogen, the organic forms of tritium are produced less readily than the organic forms of hydrogen. The organic forms of tritium are generically called organically bound tritium (OBT). Following the definitions of Kim et al. (2013), there are three types of OBT: exchangeable OBT, non-exchangeable OBT, and a special form of buried tritium (*i.e.* tritium included in the hydration shell of biomolecules). How stable tritium is within such organic compounds depends on the nature of the bond between tritium and the organic molecule and on the organic molecule affinity with the different biological tissues (ASN, 2010). When tritium is bound to oxygen, sulphur or nitrogen, it can be easily exchanged with tritium in the HTO (or $H_2O$) and the exchangeable organically bound tritium (E-OBT) is formed. When tritium is covalently bound to carbon, only enzymatic reactions can destroy the bound and non-exchangeable OBT (NE-OBT) is formed. Buried tritium, which is inaccessible because of the physical structure of the organic molecule, quickly exchanges with hydrogen atoms in the body following digestions and, consequently, it increases the amount of tritium in the body water. The time when tritium remains incorporated therefore depends on the biomolecular turnover: fast, in the case of molecules involved in the energy cycle and slow, in the case of structuring molecules or macromolecules such as DNA or energy reserve molecule. Due to longer residence in the organism, NE-OBT is of first concern for health effects of a radiological dose.

The methods to measure the HTO in environmental samples are well established but the OBT measurements are expensive and difficult. OBT measurements in all food chain components (human and non-human) are not possible. The alternative is to use radiological /environmental impact assessment models (RIA/EIA). These models can be defined as research grade and decision making models. The latter are used in design, licensing, normal operation, accident prevention and management. The decision making models must meet the following requirements:

- Relatively simple;
- Transparent;
- Easy to program;

- Results should be conservative, yet reasonable;
- Deterministic calculations possible (worst case assessments);
- Probabilistic calculations possible (95% percentile as worst case).

Finally, when the models are applied in operational context, they must quickly provide results (*i.e.* have a short run time).

During the working group meeting dedicated to "Tritium Accidents" (WG 7) of Environmental Modelling for Radiation Safety (EMRAS II) programme coordinated by International Atomic Energy Agency (IAEA) (http://wwwns.iaea.org/projects/emras/emras2/working-groups/working-group-seven.asp?s=8), one of the questions was if it is possible to have such models for tritium, due to high complexity of processes involved in the environmental tritium dynamics. A robust model for accidental tritium releases must minimize the uncertainty that can arise from the model structure, model parameters, and those situations that need special attention (see Chapter 15 in IAEA, 2014a). For routine releases of heavy water reactors, a steady state model developed by the Canadian Standard Association (CSA, 2014a) is generally used. Recently, claims have been made that the CSA model is not conservative with regard to OBT concentration in food (Thompson et al., 2015). Romania uses both Canadian (CSA, 2014a) and European practice (https://ec.europa.eu/energy/en/topics/nuclear-energy/radiation-protection) in its radiation protection programmes and some differences between those practices were noted (Galeriu et al., 2009a). The anti-nuclear groups claimed that the current models for routine release largely under-estimate the public dose because it is based on the yearly average of the air concentration and ignores spikes in the releases (Fairlie, 2010). Indeed, this claim is justified because at the receptor, the air concentration fluctuates and the equilibrium conditions are not reached as CSA (2014a) considered. In the extreme situation of a short and intense emission of tritium, a detailed analysis of the uptake processes is still needed, in order to provide a robust prediction of OBT production in crops at harvest. An analysis of the present state of knowledge regarding OBT modelling in plants during normal operation of nuclear utilities, as well as spike releases of tritium emissions are provided in the present study in order to support the current efforts regarding the tritium model improvements (Melintescu et al., 2015; Galeriu and Melintescu, 2016).

## 2. Materials and Methods

*2.1. Uncertainty of OBT/HTO ratios for routine releases*

Pressurized heavy water reactors (PHWR) were developed in Canada and have higher tritium emissions than other energetic reactors. For calculation of Derived Release Limits (DRL) for assessment of the public dose during normal operation of PHWRs, the standard guide used in Canada and Romania in the past ignored the OBT contribution to the food chain (CSA, 1987). Subsequently, the importance of OBT was pointed out in relation with fusion research (Murphy, 1993) and the consideration of OBT contribution to the food chain was proposed soon afterwards (Galeriu, 1994). The contribution of OBT was considered in the second revision of the Canadian guide (CSA, 2008). The last revision of the guide (CSA, 2014) is based on the same specific activity (SA) approach as the previous guides, assuming full equilibrium in all environmental compartments (CSA, 1987, 2008). The SA approach is also used by IAEA in its coordinated research studies (IAEA, 2009; 2010) and details regarding the CSA (2014a) and IAEA (2009, 2010) approaches are given in the Appendix A. Based on simplifying assumptions of SA approach, the ratio between OBT concentration (measured by water of combustion), $C_{OBT}$ and HTO concentration in plant water (leaves), $C_{TFWT}$ is constant. The difference between those two approaches is that CSA considers the total OBT in plant and the isotopic discrimination factor, $ID_p$ has a recommended constant value of 0.8 (range of 0.64-1.3), while IAEA considers NE-OBT in plant and the partition factor for plants, $R_p$ has a recommended constant value of 0.54 as a geometric mean (GM) and a geometric standard deviation (GSD) of 1.16. The data considered by IAEA (2009) were for barley, maize and alfalfa, grown in laboratory controlled experiments and had equilibrium values. The experimental values for the partition factor, $R_p$ cover a range of 0.4-0.68. The partition factor, $R_p$ includes both the isotopic discrimination factor (with an average value close to the $ID_p$ given by CSA (2014a)) and the contribution of E-OBT to total OBT.

For public dose assessment during routine emissions and based on equilibrium assumption, both approaches consider the crop contamination at harvest and they ignore the losses due to storage and food processing. For the assessment of plant tissue free water tritium (TFWT), CSA (2014a) considers the air HTO concentration (with a reduction factor, $RF_p$ due to the lower contribution of soil HTO), while IAEA (2009) separately considers the tritium transfer between air and plant (air pathway) and that between soil and plant (root pathway) (for details see Appendix A). The air HTO concentration is a yearly average or an average over the vegetation period. Disregarding the difference between total OBT and NE-OBT, the OBT/HTO ratio for equilibrium conditions is close to 0.7 with a range of 0.4-1.3.

Based on a detailed analysis of a large range of experimental values for OBT and HTO in agricultural products, it was observed that the ratio between OBT (water of combustion) and TFWT (HTO in leaf water) largely differs from the equilibrium value of 0.7 (CNSC, 2013; Korolevych et al., 2014; Thompson et al., 2015) with many values higher than 5 and few values higher than 10. Traditionally, this ratio is expressed as OBT/HTO, but in fact, OBT represents tritium concentration in water of combustion and HTO represents tritium concentration in plant/animal water, which is actually TFWT. Theoretically, that ratio should be expressed as OBT/TFWT, but for the sake of simplicity, the present study keeps the traditional expression of OBT/HTO ratio.

The uncertainty and under estimates of OBT concentrations come from the equilibrium assumptions based on a long term average of HTO concentration in air. In real field conditions, there is no equilibrium as it was previously pointed out (IAEA, 2008a). In fact, at the receptor where the measurements were carried out, there is a fluctuating HTO concentration in air, the HTO concentration in leaf has a diurnal variability and the OBT accumulation in different plant parts is a slow process. The sampling is usually carried out during the working hours and the HTO concentration in leaf can be low after the plume passage and high during the plume passage.

The HTO concentrations in plants reflect conditions in the few hours before sampling, while monitoring results are yearly or monthly. On the other hand, residence times for OBT in plants are sufficiently long that concentrations in these compartments better reflect average air concentrations on longer intervals. Plant OBT/HTO ratios varied between 0.12 and 0.56, when monthly HTO air average at Pickering Nuclear Power Plant (NPP) (Canada) was used (Davis et al., 2005). Modelling attempts regarding these complex situations were reported (IAEA, 2008b) and emphasised in APPENDIX A "Model Performance as a Function of Air Concentration Averaging Time" of the same study (IAEA, 2008b).

Further efforts to better link the OBT concentration to air/leaf HTO concentration were reported (Korolevych and Kim, 2013). For leaf, TFWT air concentration was averaged within 1 hour, 1 day, 3 days, and 9 days before sampling, while for NE-OBT, the averaging interval was 15 days, considering the "pod fill" period and 2 weeks before or after. Pod was a generic name for the edible part of tomatoes, red peppers, green peppers, and potatoes and fill period was 22 days after anthesis. Data for 2008, 2010 and 2011 were used to establish the reducing factors, RF for TFWT and NE-OBT and a large variability was observed between years or averaging intervals (see Tables 2 and 3 in Korolevych and Kim, 2013). Due to this large variability and huge costs of monitoring/measurements, that approach was not considered by CSA (2014a) and the "equilibrium" value of the discrimination factor, $ID_p$ was kept for total OBT.

A series of measurements with frequent sampling were carried out (Korolevych et al., 2014) assuming that the rinsed OBT is NE-OBT. A large variability of the averages of the OBT/HTO ratios up to a factor of 5 and their ranges between different years was observed (see Table 3 in Korolevych et al., 2014). The ranges of OBT/HTO ratio cover two orders of magnitude. It is useful to note that the total OBT (E-OBT + NE-OBT) is higher than the rinsed OBT (NE-OBT) by a factor less than 2 and the measured rinsed OBT can contain buried tritium. Separate experiments were carried out to assess the importance of buried tritium (Kim et al., 2008) showing that buried tritium represents less than 10 % of the rinsed OBT in vegetables. The experimental averages and ranges of OBT/HTO ratios are large and cannot be explained only by the sampling time or measurement of various OBT concentrations as it was demonstrated by a model reconstruction (see Table 4 in Korolevych et al., 2014). The model recommends a generic value of 2 for OBT/HTO ratio, but higher values were observed in certain areas (see Table 4 in Korolevych et al., 2014). Korolevych et al. (2014) concluded that the most high (and low) values of the OBT/HTO ratio arise simply as a result of normal plant sampling procedures, but SA approach does not provide an explanation for OBT/HTO range.

In a recent study (Thomson et al., 2015), concentrations of OBT and HTO were measured over two growing seasons in vegetation and soil samples obtained in the vicinity of four nuclear facilities and two background locations in Canada. For tritium processing facilities, the OBT/HTO ratios are slightly higher, with a factor of up to 2 than for CANDU reactors. The OBT/HTO ratios in vegetation have large variations of the ranges between 0.5 and 20. Ratios of the OBT activity concentration in plants to the OBT activity concentration in soils appear to be a good indicator of the long-term behaviour of tritium in soil and vegetation. The results show that some parameters used in environmental transfer models approved for regulatory assessments should be revisited to better account for the behaviour of HTO and OBT in the environment and to ensure that modelled estimates (e.g., plant OBT) are appropriately conservative. This affects not only the OBT/HTO ratios in crops, but also the HTO activity in plants in relation to the HTO activity in soil.

Following the regulatory requirements, the assessment of public dose during normal operation of nuclear facilities must be slightly conservative and based on reliable models results and experimental data. Concentrations of HTO and OBT in food items at consumption are needed. Thompson et al. (2015) pointed out few weak points of the Canadian guide (CSA, 2014a):

- There is a direct transfer between air and plants and it is supposed that the transfer to soil is only 0.3 from that in air;
- Relationships for OBT and HTO are too simple and OBT concentration as predicted is not conservative.

A recent study (Mihok et al., 2016) presents the experimental results regarding the OBT/HTO ratios in soils and plants near a gaseous tritium light source manufacturing facility in Canada (SRBT), which emits tritium as tritiated hydrogen (HT) and HTO only during the day time and week days. The plants were irrigated with tap water (containing a tritium concentration of 5 Bq $L^{-1}$), rain water (with a tritium concentration of 80 Bq $L^{-1}$) and well water (with a high tritium concentration of 11000 Bq $L^{-1}$). The plants (natural grass, sod, potatoes, beans and Swiss chard) were cultivated in barrels in soil with low tritium contamination (29 Bq $L^{-1}$ HTO and 105 Bq $L^{-1}$ OBT). The plants and soil were weekly sampled and analysed (same day and hour). When plants were irrigated with tap and rain water, the OBT concentration in soil did not vary, but the HTO concentration in soil equilibrated with that of irrigation water and with that coming from the atmospheric source (including the HT oxidation in soil). For that irrigation regime (tap and rain water), the OBT/HTO ratio in roots and tubers was about 2 and in the aerial plant parts the OBT/HTO ratio was 8. When plants were irrigated with highly contaminated water from wells, the soil and plants HTO equilibrated with about 25 % from that of irrigation water and OBT/HTO ratio in roots and tubers is 0.4 and in aerial plant parts is 1.4 (see Table 3 in Mihok et al., 2016). The experimental data show clearly that the OBT/HTO ratio is much higher than the equilibrium value of 0.7, but it is close to equilibrium when the irrigation water is highly contaminated with tritium. For tap and rain irrigation experiments, the OBT/HTO ratio increases during the development stage of plants. It is clear that the processes are complex involving photosynthesis, atmospheric and root pathways of tritium transfer and the processes are still not well understood.

*2.2. Influence of spike releases on routine emissions*

The previous experimental results clearly demonstrate that non-equilibrium processes are involved during normal operation of tritium facilities (e.g. CANDU reactors or others). Spike releases can occur due to reactor operation, for example for period of fuel change in classical reactors (Boiled Water Reactors (BWR), Power Water Reaction (PWR)) or due to technological incidents. In Pressurised Heavy Water Reactor (PHWR) as CANDU is, fuel is continuously changed and only technological incidents can produce significant spikes. Consequently, the releases are not constant and there are short periods of increased releases. An operational short term release is defined as a release which is larger than a normal release (*i.e.* higher than 2 % of 12-monthly actual or expected discharges) and occurs over a relatively short period of time (less than 1 day). For a normally uniform discharge profile, this equates to about 1 week's discharge being released in 1 day or less (IAEA, 2014b). Realistic assumptions should be used for short-term release assessments, if the annual dose exceeds 0.02 mSv (EA, 2002).

The anti-nuclear groups claimed that the public dose is largely under estimated because it is based on the yearly average of the releases, disregarding the short time higher releases (Fairlie, 2010). This claim contradicts the past experience regarding $^{131}$I and $^{137}$Cs releases into the environment stating that "the integral over infinite time of the concentration of a radioactive substance in an environmental compartment per unit of release is numerically equal to its concentration at a future steady-state when that release is repeated continuously at unit rate" (Peterson et al., 1996). In order to clarify this problem, The National Dose Assessment Working Group (NDAWG) (UK) considered in detail the case of a short-term and intense emission comparable with the same total annual emission (NDAWG, 2011). It was considered the same total emission distributed on a short period (a day or less) or distributed during a year. For spike release, three cases were analysed: a) realistic - release during a normal day over 12 hours in neutral meteorological conditions (it can include rain and moderate wind direction changes, in conformity to local multiyear data); b) cautious - release of 30 minutes in neutral meteorological condition (class D), wind direction through production area, including rain; c) cautious - 30 minutes release in stable meteorological condition (class F), wind over production area, no rain. For continuous release over a year, the average UK weather was considered. For ingestion, conservative assumptions were taken: the two foods that make the greatest contribution to dose are assumed to be consumed at the 95th percentile levels and the remaining foods at the 50th percentile levels of a distribution based on national intake rates (see Table A6 in NDAWG, 2011). The methodology for calculating the dose per unit release coming from continuous releases used the following models: ADMS 4 (version 2008) (CERC, 2008) and PC CREAM (version 1997) (Mayall et al., 1997), while for spike release, the following models were used: FARMLAND (version 1995) (Brown and Simmonds, 1995) and SPADE (version 1999) (Mitchell, 1999). In parallel, for tritium, TRIF (acronym for TRitium.transfer Into Food) model (Higgins, 1997) was used.

The considered tritium release is 1 TBq for the whole year (continuous) or for the spike release. For the realistic case (atmospheric stability of class D, 12 hours) the wind direction is not uniformly distributed because this case has a very low probability, but it can vary according to local statistics considering successive hours and the same stability class D. The detailed results for each radionuclide of interest and case considered (infant, child, adult) are given (see Tables A10-A21 in NDAWG, 2011) and for tritium, the results are given in Table 1 (present study). In Table 1, it is observed that a short-term emission determines an ingestion dose much higher than that of the same emission distributed during a year and in case of category F of atmospheric stability, the ingestion dose is with a factor of about 35 higher than that in the case of a continuous release. The general conclusion in the report was: "Where there are only annual limits in place, **and it is cautiously assumed that discharges occur at these limits over a short period of time**, then doses from the assessment of a **single realistic** short-term release are a factor of about 20 greater than doses from the continuous release assessment" (NDAWG, 2011).

*2.3. Uncertainty of tritium dynamics for short term atmospheric exposures*

For OBT production in crops following a short atmospheric HTO exposure, a review of experimental data was undertaken (Galeriu et al., 2013) and various data for OBT concentration in different plants were compared based on the translocation index (TLI). TLI is defined as the percentage of OBT concentration in plant part (combustion water) at harvest related to the HTO concentration in leaf at the end of exposure. With few exceptions (cherry tomato and tangerine), TLI at night is lower than that at day. It was also observed that the transfer rate between air and leaf at night is 2-10 times lower than that at day emphasising that stomata are not fully closed at night.

In WG 7 (Tritium Accidents) in the frame of IAEA EMRAS II programme, the problem of OBT formation in crops was discussed based on experimental data for winter wheat used in a models testing exercise (see Chapter 9 in IAEA, 2014a). The experiments were carried out in Germany in 1995-1996 at Kernforschungszentrum Karlsruhe (KfK now KIT) where winter wheat plants were grown in a small experimental field and short-term (one hour) HTO exposure experiments were conducted during the grain-filling stage of the wheat, at different hours of the day. The details regarding the experiments were given elsewhere (Strack et al., 1998; 2005; 2011; IAEA, 2014a). It is important to note that the soil was isolated and has very low HTO concentration and that experiment considers only the atmospheric pathway. The experimental data in wheat scenario and the analysis of models predictions were partially published (IAEA, 2014a). A detailed analysis of the data and models is presented in the present study emphasising the processes of OBT formation at day and night (see Results and Discussion).

## 3. Results and Discussion

### 3.1. Uncertainty of OBT/HTO ratios for routine releases

In field conditions the equilibrium conditions are not reached and consequently, the dynamic models need detailed dynamics of HTO concentration in air. The measurements of HTO concentration in air at an hourly time step at a certain distance from a nuclear facility are very difficult and time consuming involving large financial resources. Consequently, various attempts were carried out to reconstruct the air HTO concentration from the measurements on stack emission, meteorological data and data for other radionuclides such as $^{85}$Kr for fuel reprocessing plants (Maro et al., 2016) and $^{41}$Ar for Nuclear Research Unit (NRU) (Korolevych and Kim, 2013; Korolevych et al., 2014) emitted together with tritium. The reconstruction of the air HTO dynamics is not a part of the present study. In the present study, a simulation of the dynamics of HTO concentration in air is undertaken for the CANDU reactors at Cernavoda NPP (Romania), in order to emphasise the fluctuations of the HTO air concentration and to point out the processes affecting the OBT/HTO ratio. In Romania operating two CANDU reactors, the atmospheric emission is monitored daily at both units starting from 2009. The simulated HTO air concentration is based on the daily measured emissions of both units in 2011 and on the hourly meteorological data base of the utility. Simulations of the dynamics of HTO concentration in air at various receptors were carried out, based on the atmospheric transport model ISC PRIME (Schulman et al., 1997) including the building effects, details on building dimensions and positions, and elevation of the receptor (Cernavoda is a complex site, not a flat horizontal case). For a receptor outside of the exclusion zone at 1.7 km in NW direction, the simulation of the hourly HTO concentration in air is given in Fig. 1 and large fluctuations can be observed. For this simulated example, the yearly average HTO concentration in air is 0.9 Bq m$^{-3}$. In fact, a series of spikes with amplitudes of up to 100 times larger than the yearly average occurred. If the diurnal distribution (during 24 hours) is considered, the average HTO concentrations in air during the day time of 0.52 Bq m$^{-3}$ is significantly lower than those during the night time (Fig. 2). It is interesting to observe that for this particular simulation (Fig. 2) the average of air HTO concentration for evening to midnight (of about 1.8 Bq m$^{-3}$) is higher than that for midnight to sun rise (of about 0.98 Bq m$^{-3}$). The simulations were carried out for other receptors positions showing a large variability of day/night average, but systematically, the number of hours without plume is much bigger than those with plume above the receptor. The measurements on soil HTO at the monitoring station outside of the exclusion zone show that HTO concentration in soil water are 10-60 Bq L$^{-1}$ in the upper 20 cm of soil (Bucur, C, personal communication, 2016) and most of the time, the leaf water HTO concentration depends on soil HTO concentration through plant transpiration. Transpiration rate has a diurnal pattern and can be important at night (Caird et al., 2007). Crops have various root lengths during their development stages and the depth of water extraction in soil can be larger than 1 m. The HTO concentration in root soil slowly varies, but the transpiration flux has a strong diurnal pattern and finally, the HTO concentration in leaves varies significantly during the day and night time. For fluctuating HTO concentrations in air, the transfer rate between leaf and air is less than 1 h$^{-1}$ during the day time and it is 2-10 times lower during the night time (Galeriu et al., 2013). Consequently, the models must consider the day and night transpiration and characteristics of root uptake in more details, in order to better assess the leaf TFWT and OBT dynamics, because the dynamics of leaf TFWT directly affects the OBT formation rate and accumulation. In fact, the OBT concentration in crops at harvest is needed for the assessment of public dose and OBT concentration is affected by the history of the crop contamination during its whole development period and mostly, in the last 1-2 months. For the same leaf HTO concentration, the OBT formation during the night time represents 1/3-1/10 from that during the day time (Galeriu et al., 2013) and in Fig. 2, it can be observed that night air HTO concentration is higher than that at day. If a plant (or genotype of plant) has a relative high transfer rate between air and leaf (and/or high night transpiration) and the air HTO concentration at night is 2-5 times higher than that at day, it is possible that OBT production at night to be comparable with that at day. The OBT production at night was previously pointed out as a topic of interest for modelling purposes (IAEA 2008a; IAEA 2014a), but in practice, only UFOTRI (Raskob et al., 1996) and IFIN (Galeriu et al., 2000a) models considered it using calibration with experimental data for spring wheat. That calibration was useful for winter wheat and rice and in some limits, for a genotype of soybean (Raskob, 2007). There are many crops of interest and recent experimental data (Shen and Liu, 2016) shows that for at least soybean, during night time conditions, the OBT formation cannot be ignored. Only recently, a model considering OBT production at night without any calibration with experimental data was proposed (Galeriu and Melintescu, 2016) but further studies are still needed, in order to cover various crop types and finally, tests with well-designed experiments.

Figs. 1 and 2 are based on simulation of air HTO concentration at an hourly time step and have illustrative purposes. Many simulations were carried out for different receptors directions and many years and the importance of root uptake (plume off) and night OBT formation remain valid.

In the previous paragraphs, the importance of transpiration and night OBT production in field conditions with fluctuating air concentration on OBT/HTO ratio were pointed out. Another issue needs a further discussion: when and what has to be measured. The interest resides in crops at harvest and consequently, the plant sampling at an early development stage is not relevant. If the sampling is done when the plant is relatively mature, significant quantity of OBT is accumulated until sampling. Sampling before plume arrival results in low TFWT and high OBT stored before the plume arrival and consequently, high OBT/HTO ratio is obtained. Sampling during the plume or immediately after the maximum of the plume, results in high TFWT and high new OBT production, added to the OBT stored before the plume arrival. The new OBT, compared to the previously stored OBT, is only a fraction and in fact, OBT/HTO ratio is relatively low. It is also important what has to be measured: total OBT or NE-OBT. Interesting results were recently reported (Kim and Korolevych, 2013). For crops with high carbohydrate content, the experiments were carried out in a polluted area and after a first sampling, the crops were irrigated with water containing high HTO concentration (treated case). Weekly average of air HTO was measured and samples were analyzed for total OBT and NE-OBT. In all treaded crops, E-OBT fraction was higher than for the untreated case and for treated and untreated cases, the differences between crops were significant (reflecting crop biomass composition). On average, the E-OBT fraction was 34 % for untreated and 44 % for treated crops. These values are close to past results regarding the fraction of unbound H in carbohydrate, but are lower for crops with high protein contents (grains of cereals) or lipid content (sunflower) (Diabate and Strack, 1993). The HTO concentration in irrigation water was about 20 times higher than the average air concentration and significantly increased the total OBT at sampling time, but marginally increased the E-OBT fraction.

It may be concluded that NE-OBT is a variable fraction from total OBT and depends on the plant type, irrigation conditions and sample treatment for OBT measurements. As the data base for the observed OBT/HTO contains total OBT or NE-OBT, this variability can partly explain the large range of observations, far from the equilibrium value. The main factors of variability remain the fast dynamics of TFWT and the OBT formation and its slow accumulation.

The previous analysis of OBT/HTO ratio in environmental products (Korolevych et al., 2014; Thompson et al, 2015) were bulked together for all types of environmental samples taken from various sites. In the present study, the experimental data for HTO and OBT concentrations in various environmental samples taken from Wolsong NPP (Korea) (KEPCO, 2012; 2014) were analysed for the same sample type and site. The OBT/HTO ratios were separately analysed for grains (barley and rice), vegetables (cabbage and persimmon) and milk, in order to see if a dependence of the OBT/HTO ratio on HTO concentration in sample occurs. It is observed that for grains (barley and rice), the average OBT/HTO ratio is about 2, slightly higher for barley than for rice with quite large spread of data (Fig. 3). The OBT/HTO ratio for cabbage is close to 0.45 and for persimmon to 0.58 with little spread of data (Fig. 4). It is observed that the OBT/HTO ratio for cabbage and persimmon are close to the expected equilibrium value of 0.7, but for grains is 2-3 times higher. There are no data for grass, but the expected OBT/HTO ratio for milk is 0.25 – 0.4, depending on the contamination of the animal drinking water and based on the equilibrium assumption of CSA (2014a). The average OBT/HTO ratio for milk is 1.2 with a moderate spread of data (Fig. 5) and it is 3 – 5 times higher than the expected one. The results in Figs. 3 – 5 pointed out that the OBT/HTO ratio depends on the crop type and receptor location.

From the previous discussion, it results that OBT/HTO ratio largely depends on receptor location, plant type, and history of the dynamics of HTO in air. Consequently, it is very difficult to model these complex processes and the models need major improvements and cannot be simple. The nuclear regulatory bodies need simple models, slightly conservative for public dose assessment and the CSA methodology (CSA, 2014a) was criticised not to be conservative (Thompson et al., 2015). In the present study, a probabilistic application of CSA model (2014a) is proposed with parameters having a probability density distribution, as it was recently reported but not for tritium (Simon-Cornu et al., 2015).

The approach regarding the tritium transfer parameters taken by the CSA (2014a) is to derive the pond or well water based on the ratio between soil water and air moisture of tritium content, $RF_{sw}$, with a default value of 0.3. The uncertainties in the $RF_{sw}$ values can largely contribute to dose uncertainty (about 28 %). A larger set of experimental values from Russia, France, Canada, India, and Japan was used elsewhere (IAEA, 2009) having a log-normal distribution with a geometric mean (GM) of 0.23 and a geometric standard deviation (GSD) of 1.7 (see Table 1 from Miscellaneous Topics in IAEA (2009)). The data in that table suggest that southern regions (e.g. France) or humid regions (e.g. Japan) may have higher values of the ratio of soil water to that of air moisture concentration. Based on these values, a default value of 0.3 is reasonable and consistent with the older recommendation of IAEA (2003). A value of 0.5 is likely to be conservative, although values as high as 1.0 are possible. Consequently, experimental values based on local measurements should be used whenever is possible or selected from general information on climate and precipitation.

Tritium transfer from air to plants in the CSA study (2014a) uses the reduction factor, $RF_p$ (see equation (A2) in Annex A) with a default value of 0.68, but it has a large variability in practice. Based on the IAEA study (2009), the range of the reduction factor, $RF_p$ can be assessed considering the variability of the relative humidity, RH (see equation (A5) in Annex A) and the ratio between tritium concentration in soil water and air moisture. Combining equations (A2) and (A5) from Annex A, the dependence of the reduction factor, $RF_p$ on both the relative humidity, RH and the ratio between tritium concentration in soil water and air moisture, $C_{sw}/C_{am}$ is obtained (see equation (1)):

$$RF_p = \frac{RH + (1-RH)*\frac{C_{sw}}{C_{am}}}{\gamma} \qquad (1)$$

The ratio between tritium concentration in soil water and air moisture, $C_{sw}/C_{am}$, as well as the ratio between TFWT in plants and air moisture, $C_{TFWT}/C_{am}$ for various values of relative humidity, RH is given in Table 2. The bolded values in Table 2 represent the reduction factor, $RF_p$ (in conformity with equation (1)). In Table 2, large differences for reduction factor, $RF_p$ are observed. For humid areas, the soil water is affected by rain. The scavenging ratio (*i.e.* the ratio between tritium concentration in rain water and that in surface air moisture) varies between 0.1 and 10 (Ota and Nagai, 2012) and rain affects the soil water concentration. The scavenging ratio depends on the distance between the emission stack and the receptor and the effective release height. For CANDU 6 reactors with low stack height, the HTO concentration in soil water is affected mostly by the HTO concentration in air moisture. Consequently, the plant water is site specific, not generic. In the absence of relevant local data, a probabilistic approach can be taken and for the reduction factor, $RF_p$ an average of 0.8 with a SD of 0.1 can be provided.

The values of the other parameters involved in equation (A3) in Annex A (dry matter fraction of plant, $DW_p$, isotopic discrimination factor, $ID_p$, and water equivalent of organic matter, $WE_p$) vary with the plant type. For example, based on experimental data, the dry matter fraction of terrestrial plants, $DW_p$ has a log-normal distribution with various GMs and GSDs depending on plant type (Table 3). The water equivalents factor, $WE_p$ depends on plant composition and has close GMs and GSDs for various plants (Table 4). The water equivalent factor, $WE_p$ can be bulked for all plants with a GM of 0.51 and a GSD of 1.1 (see Table 63 in IAEA (2010)).

For equilibrium assumptions, the isotopic discrimination factor, $ID_p$ (see equations (A1) and (A3) in Annex A) has a range between 0.64 and 1.3 (CSA, 2014a) and a default value of 0.7 is used. It is implicitly assumed that the OBT concentration at sampling time depends on a long term average of HTO concentration in crops. The equilibrium between HTO concentrations in all environmental compartments (air, soil, leaves) is never reached in real field conditions. When the experimental data are used, the ratio between OBT in combustion water and HTO in plant water varies and a generic average value of 2 is recommended (Korolevych et al., 2014), even certain data sets indicate higher values. Based on the previous discussion regarding the OBT/HTO ratio, the isotopic discrimination factor, $ID_p$ is rarely close to the equilibrium value, depending on crop, receptor location, and the detailed dynamics of HTO concentration in air. In the absence of site specific values of OBT concentrations in crops, the present study proposes to consider a log-normal distribution for the discrimination factor, $ID_p$ (= $R_p$) with a minimum value of 0.3, a maximum value of. 10, a GM of 1.732, a GSD of 1.77, and a CV of 0.62. The average value is close to the recommended value of 2 (Korolevich et al., 2014), but the maximum value allows a conservative estimate.

In order to emphasise the importance of parameters uncertainty, in the present study the case of Cernavoda NPP (Romania) is considered, based on a conservative diet and significant local food consumption. The yearly average concentration of HTO in air at the location of interest is assumed to be 1 Bq m$^{-3}$. Several scenarios based on the age groups proposed by ICRP (2006) were considered (infant, child and adult):

- OBT/HTO ratio has the equilibrium value of 0.7 as in CSA (2014a) and the tritium concentration in the animal drinking water is the same with that of drinking water for humans (4.5 Bq L$^{-1}$). The results are given in Table 5.

- OBT/HTO ratio has the equilibrium value of 0.7 as in CSA (2014a) and the tritium concentration in the animal drinking water is 9 times higher than that of drinking water for humans (*i.e.* pond water with a tritium concentration of 40 Bq L$^{-1}$)., close with the conservative recommendation in CSA. The results are given in Table 6.

- OBT/HTO ratio has a conservative value of 10 and the tritium concentration in the animal drinking water is the same with that of drinking water for humans (4.5 Bq L$^{-1}$). The results are given in Table 7.

In Tables 5-7, it can be observed that the uncertainty introduced by the OBT/HTO ratio is comparable with that introduced by tritium concentration in animals' drinking water and the conservative dose for public is only 2 µSv year$^{-1}$ for a yearly average concentration of HTO in air of 1 Bq m$^{-3}$.

In all three scenarios the food production was maximised, as well as food intake. In fact, it is important to have site specific food production (16 sectors and radius of 1-5 km) and the consumption of local food. Based on site specific information, the exaggerate conservatism is avoided.

In a PHWR as CANDU is, 97 % of atmospheric emission is HTO, during the day time and night time. A CANDU reactor differs from the SRBT case (Mihok et al., 2016) where the emissions are predominately as HT, restricted to working hours (7.00 - 17.00), and stopped during the weekends. There is no such a facility in Romania, but some experience for detritiation facility and other tritium bearing utility exists. There are some peculiarities at SRBT which potentially can explain the results of Mihok et al. (2016) and each of them will be shortly discussed in the following. The experimental errors can be eliminated because special quality assurance was taken (St-Amant, 2016). HT is very little oxidised in contact with the air or leaves (Ichimasa et al., 1999) but strongly oxidised in soil (Ichimasa et al., 1999; Ota et al., 2007) enhancing the HTO concentration in root soil. The increased HTO concentration in the transpiration flow due to HT oxidation in soil is moderate and cannot explain the experimental data provided by Mihok et al. (2016). For tritium handling facilities, various species of soluble tritiated organics were reported (Belot et al., 1993; 1995) and tritiated formaldehyde can be incorporated in leaves increasing the OBT concentration (Belot et al., 1992). Tritiated methane is also a potential substance contributing to leaf OBT (Amano, 1995; Kakiuchi et al., 2002). It seems that at SRBT, the emissions of other tritiated substances than HT and HTO have a very low probability, but cannot be fully excluded (MacDonald, J, personal communication, 2016) and more accurate measurement can be carried out in the future for clarifying this potential cause. The soil around SRBT has a high concentration of OBT due to high tritium emissions occurred in the past and a rain splash process can contaminate the plants (Dunne et al., 2010). The rain amount and intensity was low during the experiment (Clark et al., 2014) and rain splash is not enough to explain the results. If no other impurities excepting HT and HTO were emitted, the explanation might be the fact that the

OBT formation in crops is not yet well understood and the present models have large uncertainties for OBT predictions.

*3.2. Influence of spike releases in routine emissions*

The results regarding tritium contamination of foodstuff in case of spike releases considered in the study of NDAWG (2011) are based only on UK standard models. In parallel, TRIF model was investigated. TRIF model (Higgins, 1997) has many compartments with various transfer rates fixed at generic values. It considers only NE-OBT and assumes that there is a loss of plant OBT to HTO with a half time of 10 days. The foodstuffs considered in TRIF model are green and root vegetables, fruits, cow milk, cow meat and liver, sheep meat and liver. TRIF model does not consider the cereals.

The results of NDAWG study (2011) for realistic case (class D, 12 hours) can be compared with TRIF model considering integrated food contamination for a similar deposition of tritium of 1240 Bq m$^{-2}$ and are given in Table 8. In Table 8, it can be observed that NDAWG results (2011) are about 2 times more conservative than TRIF model results (Higgins, 1997). When TRIF model is compared with SA model for routine release, it gives an activity about 30 % higher in food products. NDAWG (2011) finally selected the most conservative results.

The routine releases were considered as a special task in the frame of an IAEA coordinated research programme: Biosphere Modelling and Assessment (BIOMASS) (IAEA, 2003). Based on data from Canada, France and Russia, BIOMASS programme provided inter-comparison results between models and observations, including OBT in plants, following chronic atmospheric releases (IAEA, 2003). The final BIOMASS report (IAEA, 2003) concluded that:

- The conceptual model of OBT formation may need to be rethought for sites that are subject to a fluctuating airborne plume. None of the models was able to simulate the observed increase in plant OBT concentration with time;
- Unlike steady state releases, a non-planned release detected at the stack may not have an effect on all the media compartments.

A model inter-comparison exercise was carried out for a hypothetical tritium accident with one hour emission of 3700 TBq (Guetat and Patryl, 2008). The atmospheric transport was imposed to all participants and the radiological doses coming from food ingestion varied on a large range for class D (with rain) and class F (night). If the release is decreased to 1 TB, the radiological doses have a range between 0.27 and 3.4 µSv year$^{-1}$ for cautious class D and a range between 1.1 and 17 µSv year$^{-1}$ for cautious class F. The values in Table 1 are in these ranges but close to the maximum values.

In the present study, the same conditions as in NDAWG (2011) were considered, but different models for spike or continuous emissions were used. For spike releases, the UFOTRI model (Raskob, 1993) and an upgraded version of IFIN-HH model (Melintescu and Galeriu, 2005) were used but with a time step of one hour. The IFIN model (Melintescu and Galeriu, 2005) is a partially improved version of the tritium module developed in the frame of Real time On-line DecissiOn Support System (RODOS) project (Galeriu et al., 2000a; 2000b). For a normal day case (class D, 12 hours), the hourly meteorological data from Cernavoda NPP (Romania) were used and a case where wind changes directions up to $150^0$ was selected. For cautious cases (class D and F), the wind direction was chosen to cover the sector with the maximum food production. Recommendations of CSA for accidental releases (see sub-chapters 5.3.3. and 5.3.4 in CSA, 2014b) were followed. For ingestion, as in NDAWG (2011), the most important food items (milk and vegetables) were considered at 95 % from the local intake distribution data and the local consumption was maximised. For continuous release, CSA model (2014a) was used (in the probabilistic approach proposed with conservative assumptions), as well as NORMTRI model (version 1999) (Raskob, 1994).

In Table 9, it can see that the occurrence of spikes in routine emissions increases the public dose, but predictions cover a large range, reflecting the assumptions considered by the models. In USA, for routine emissions of tritium, CAP88-PC model is used and gives a public dose 3 times higher than NORMTRI model (Imboden and Overcamp, 2005). All models assessing the public dose needs the air HTO concentration, in order to derive HTO and OBT in crops at harvest and the atmospheric transport calculations are subject to large uncertainty. Comparing the results of CAP88-PC, NORMTRI, and ISC-PRIME models for atmospheric transfer with the measured yearly average of air HTO concentration, those models give good predictions, but for monthly or biweekly average of air HTO concentration, large miss-predictions are observed (Michelotti et al., 2013; Galeriu and Melintescu, 2012). As it was previously discussed in the present study, plant HTO and OBT concentrations depend on air HTO concentration during the previous day (HTO) or month (OBT). Based on the previous considerations, it is clear that the OBT dynamics in plants is a complex process and equilibrium conditions are never attained in practice. Plants used for human consumption have different harvest times and development stages. A spike release can occur any time during the year and crop contamination depends on its development stage at the day of release. For Romanian diet and crops near Cernavoda NPP (Romania), the day of release was varied at each mid-month and the short and intense emission was supposed to occur near mid-day. The IFIN tritium model (Melintescu and Galeriu, 2005) was used in that case and a significant seasonal pattern of the public dose was observed with a maximum value in September (when maize is harvested) and a minimum value in winter (Fig. 6). Local habits regarding food and feed are very important and can vary in each country, as well as the share of the local production in the diet. The Romanian results must be considered as an illustrative example only.

In Romania, the atmospheric emission of both CANDU reactors is about 400 TBq $y^{-1}$ in 2011 - 2015. The HT and HTO releases at stack are monitored daily and monthly values of HTO concentrations in air, water and food are measured at various receptors. The HT emission represents 1-3 % of the total emission. Public dose is assessed in two ways: using conservative estimate of dispersion factors and Canadian standard (CSA, 2014a) or using only monitoring results. When the CSA (2014a) is used, the public dose is higher than when the monitoring results are used, but never exceeds 10 µSv $y^{-1}$. For a yearly emission of 1 TBq, the public dose is 0.025 µSv $year^{-1}$, a value much lower than those in Table 9 (fifth column). For all period from 2009 until present, a single significant spike was detected (6 times higher than the weekly average), but it did not occur in the vegetation period and the dose for public marginally increased.

The assertion of NADWG (2011) that the public dose from a short term release is much higher than those from the same amount emitted but for the whole year, must be reanalysed considering the high uncertainty of the atmospheric transport for a short term (hour, week), the non-equilibrium processes in field conditions, the mobility of tritium, the fast tritium transfer from air to leaf, and the slow accumulation of OBT, as well as the tritium recycling. Routine models considering equilibrium conditions and simple atmospheric transport models involve large uncertainty. The tritium reemission from soil and plant, as well as respiration process, produce a secondary tritium plume. The OBT accumulated in soil decomposes slowly and is recycled, involving that the processes must be modelled on large space and time domains (but with fine time steps and space grid) and moreover, that all processes are well understood. The uncertainty due to OBT production in crops is only a part of the problem.

Due to the insufficient understanding of processes involved in tritium transfer in environment, it is not possible to assess the public dose with low uncertainty and moderate conservatism. Considering the previous results and analysing the present capability of the models to predict the consequences of routine or spike release, it is obvious that further studies are needed in order to clarify if tritium transfer in environment differs from other radionuclides transfer like iodine and caesium and a spike release of tritium gives higher dose than models predict.

*3.3. Uncertainty of tritium dynamics for short term atmospheric exposures*

In wheat scenario (Chapter 9 in IAEA, 2014a), the experimental data on HTO and OBT concentrations in leaves and ears (grains) were used for a series of one hour exposure in an experimental chamber, at different hours of the day. During the exposure, the air HTO concentration, temperature, light and relative humidity were measured. At the end of exposure, the chamber was open, and samples of leaves and ears were taken immediately, but also on the subsequent hours including the normal harvest time (Strack et al., 1998; 2005; 2011). The HTO and NE-OBT data were organised in a data base for each experiment containing the start of one hour exposure, HTO and OBT in leaves and ears at the end of exposure, and the subsequent samplings, including the last one at normal harvest. The data base contains 14 experiments starting at 7.00 in the morning and ending at 23.00 in the evening (denominated as s7, s8, s10, s11, s20p, s23p). In the present study, based on that data base, the OBT and HTO concentrations in leaves at each sampling were analysed in order to assess the dynamics of OBT/HTO ratio in the first 24 hours after the end of exposure (Fig. 7). In Fig. 7, it can be observed that the OBT/HTO ratio has very low values at the end of each exposure and gradually increases up to a value of 10 after 24 hours, demonstrating that the process of OBT formation is complex and not fast and the OBT/HTO ratio strongly depends on sampling time after exposure.

In wheat scenario (IAEA, 2014a), the initial conditions for the one hour exposure were used and the modellers were asked to predict, among others, the OBT concentration in grain (ear) at the end of exposure and its dynamics for the next 24 hours and at the harvest (Bq kg$^{-1}$).

Four countries/models, participated at this scenario:
- JAEA (Japan Atomic Energy Agency, Japan) (with SOLVEG model ), a complex research grade model (Ota and Nagai, 2011);
- CEA (Commissariat of Atomic Energy, France) (with CERES model), a simple model with a constant exchange velocity for day or night and with OBT production depending on the integrated leaf HTO concentration (Patryl and Armand, 2005; 2007);
- IFIN ("Horia Hulubei" National Institute for Physics and Nuclear Engineering, Romania) (with FDMH+ model), a research grade and operational model of moderate complexity (Galeriu et al., 2000a; 2000b; Melintescu et al., 2002; Melintescu and Galeriu, 2005), with an intermediate upgraded version (Galeriu et al., 2009b; Melintescu and Galeriu, 2011);

- KIT (Karlsruhe Institute of Technology, Germany) (with UFOTRI model), a research grade and operational model of moderate complexity (Raskob, 1993), with an intermediate upgraded version (PLANT-OBT) (Raskob et al., 1996; Raskob, 2011).

The OBT concentration in grain at harvest is the most important endpoint and modellers were asked to submit predictions for many observations and time steps, in order to detect the potential compensatory errors and to help the models improvements. For grain at harvest the predicted to observed ratio (P/O) varies between models (Fig. 8).

At harvest, the JAEA model exhibited a large spread in its P/O ratios for grain OBT concentrations with practically no OBT in grain predicted at night. The KIT model over-estimated most observations, while the other models (CEA and IFIN) mostly under-estimated. The models predictions are acceptable for that exercise (see Chapter 9 in IAEA, 2014a). For leaf OBT concentration at the end of exposure, KIT model is close to experimental data, but CEA and IFIN models under estimate the experimental data by a factor of 2-10 (Fig. 9). JAEA model gives good predictions excepting the night case (Fig. 9).

For leaf OBT at harvest, the models predictions have large variations and any models results are not able to satisfy the statistical criteria for an acceptable performance (Fig. 10). The IFIN model does not predict practically any OBT concentration in leaf at harvest and consequently, the model must be improved for a better understanding of processes.

The experimental data on leaf and ear HTO concentration and grain OBT concentration at harvest were analysed (Strack et al., 2011) and an empirical relationship between OBT concentration in grain at harvest (water of combustion) and integrated HTO concentration in leaves and ears was found distinguishing between OBT produced during the day and night time (Strack et al., 2011):

$$C_{OBT-grain} = 0.48 * [IC_{leaf-day} + 0.2 * IC_{leaf-night} + 0.5 * (IC_{ear-day} + 0.2 * IC_{ear-night})] \qquad (2)$$

where: $C_{OBT-grain}$ is OBT concentration in grain at harvest (Bq mL$^{-1}$); $IC_{leaf-day}$, $IC_{leaf-night}$, $IC_{ear-day}$, $IC_{ear-night}$ are HTO integrated concentrations in leaf and ear during the day and night time, respectively (kBq h mL$^{-1}$); 0.48 is a regression coefficient.

The equation (2) indicates that the ear contributes with a half to the overall OBT production in grain at harvest and the OBT produced during the night time is 0.2 of that produced during the day time. The constant slope in equation (2), 0.48 can be justified for the experimental conditions where the grain filling is linear. The value of the slope itself, 0.48 is a characteristic of the German wheat cultivar. It is well known that for cereals, the ear contributes to the photosynthesis in the period before the yellow ripe (Goudriaan and van Laar, 1994) and the 0.5 value in equation (2) is an average for the period considered in the experiment (days 17 – 28 after anthesis and the harvest was 47 days after anthesis). The equation (2) clearly demonstrates the importance of the HTO concentration in leaves (and ears for cereals) for the OBT production.

All the models participated to wheat scenario consider the dependence of OBT production on leaf HTO concentration but based on different approaches. In the complex model SOLVEG (Ota and Nagai, 2011) the OBT is produced in leaves in a carbohydrate sub-model depending on leaf HTO concentration and photosynthesis rate. The OBT production rate depends on plant type and growing stage, also. The SOLVEG underestimation of OBT production at night could be due either to very low stomatal conductance at night or to carbohydrate sub-model. The KIT model (PLANT-OBT) (Raskob et al., 1996; Raskob, 2011) considers the dependence of OBT production on HTO concentration in leaf and ear but the model needs too many parameters calibrated for crop type (wheat). The standard KIT model (UFOTRI) (Raskob, 1993) considers the dependence of OBT production on only HTO concentration in leaf, plant type and development stage and includes separately the OBT production at night calibrated with experimental data. UFOTRI was successfully tested for other crops as rice and soybean (Raskob, 2007). The IFIN model (Galeriu et al., 2009b; Melintescu and Galeriu, 2011) considers the dependence of OBT production on HTO concentration in leaf but indirectly considers ear in photosynthesis increasing the leaf area index (LAI). IFIN model as UFOTRI uses calibration for OBT night production. IFIN model considers fast and slow maintenance respiration, dry matter and OBT partition for different plant parts, but largely under-estimates the leaf OBT concentration at harvest. The simple CEA model (CERES) (Patryl and Armand, 2005; 2007) considers a linear dependence of OBT concentration in grains on the integrated HTO concentration in leaf. The slope of that dependence can be adjusted by the duration of crop growth. The CEA model does not consider the crop type and the development stage. It considers a prescribed value of leaf resistance for day (300 s m$^{-1}$) and night (3000 s m$^{-1}$). For wheat scenario, CEA model considers the duration of grain growth and not the duration of plant growth. The model is not documented enough in order to explain the OBT dynamics in leaves for cereals.

It is difficult to clarify which model depending on its level of complexity better predicts an accidental situation with tritium emission, based only on the experimental data sets for a single crop grown in specific environmental conditions. Consequently, many tests are necessary for model selection covering many plant types and various environmental conditions.

The present study analyses the experimental results regarding NE-OBT but in many cases, the measurements consider total OBT. Considering the field conditions at Perch Lake plot garden of Atomic Energy of Canada Limited (AECL) (Kim and Korolevych, 2013), the ratio between NE-OBT and total OBT of various vegetables (cabbage, lettuce, tomato, radish, and beet) varies between 0.46 and 0.74. There are no data immediately after an acute exposure of wheat for total OBT and the results in Fig. 7 emphasise that the formation of NE-OBT is a long process starting with low values of OBT/HTO ratio. A lower value of NE-OBT/Total-OBT ratio can be suspected immediately after the plume passage. Further dedicated controlled experiments would be necessary.

## 4. Conclusions

The variability of OBT/HTO ratio on a large range is a result of no equilibrium conditions in real field situations and of the difference between the HTO and OBT dynamics in crops. Plants used for human consumption have different harvest times and development stages. The OBT/HTO ratio at sampling time depends on the tritium dynamics at stack emission, atmospheric transport at receptor location and crop type. A general dependence of OBT/HTO ratio on HTO concentration in crops cannot be observed for experimental data at Wolsong NPP (Korea), but trends for each crop type and receptor position (KEPCO, 2012; 2014) are foreseen. The Canadian Standard (CSA, 2014a) assuming equilibrium conditions can be used but in a probabilistic approach or in a deterministic case with an increased value of isotopic discrimination factor, $ID_p$. For routine releases, the public dose is still very low (with realistic local habits and food production) and the uncertainties of the model predictions have no consequences on any health effects.

The SRBT case (Mihok et al., 2016) is not specific to conditions of a CANDU reactor and can be affected by the emission of organic tritiated species (to be tested), and the uncertainty of the current models (still not crop specific).

The spike release influence on the yearly public dose cannot be ignored and is due to non-equilibrium processes in real field conditions. If the spike coincides with the period when most of the crops growing around the nuclear site are about one month before harvest, the food contamination for human consumption can be higher than in the case of a uniform emission (the same emission as in the spike release) during a year. In temperate climate, there is no plant growth in winter. The uncertainty of the present dynamic models does not allow a quantitative assessment.

The analysis of wheat experiments (Strack et al., 1998; 2005; 2011) for a short term but intense atmospheric emission emphasises that the OBT formation is a long process (about 24 hours), there are some differences between night and day releases and the leaf (and ear) HTO dynamics is more important than it was previously assumed. That experiment considers the cereals in their linear grain filling period and the models results cannot be generalised for other crops, but point out the models limitations. The requirements for a robust operational model (see Introduction) are not accomplished by any model participated to wheat scenario. Due to the complexity of processes involved in OBT formation and HTO dynamics, further studies are still needed. The efforts must be oriented to a better understanding of the root and air pathway balance on HTO dynamics subject to various environmental conditions and to the potential role of OBT production at night. Further studies highly depend on each country nuclear energy policy, nuclear regulatory body requirements on model uncertainty and mostly, on the available resource. A large international collaboration is needed, directly between various teams or coordinated by IAEA as practiced from many years.

**Acknowledgements**

This study is part of the project EXPLORATORY IDEAS 65/2011 financed by the Romanian Authority for Scientific Research. The excellent collaboration with the Radioprotection team of CANDU NPP (Romania) is highly appreciated. The past and present exchange of ideas with WG 7 members of EMRAS II programme is acknowledged. We thank to Dr. Siegfried Strack (Karlsruhe Institute of Technology, Germany, retired) for providing us full access to his data base on wheat experiments. Last but not least, we want to thank to Dr. Sylvie-Ring Peterson (Lawrence Livermore National Laboratory, USA, retired) and Dr. Gretchen Gallegos

**Appendix A**

*CSA and IAEA guidance for tritium in crops*

The assessment of public dose following an atmospheric routine release of $^{14}$C, $^{36}$Cl and $^{3}$H was carried out by IAEA in its coordinated research studies (IAEA 2009; 2010) based on specific activity (SA) approach used for equilibrium conditions. SA is defined as the radionuclide activity per mass of the stable element. The SA approach for equilibrium conditions is also used by CSA (2014a).

CSA (2014a) proposed the following equations for OBT and tissue free water tritium (TFWT) concentrations in plants on a fresh weight basis:

$$C_{OBT} = \frac{RF_p * DW_p * ID_p * WE_p * C_{air}}{H_a} \tag{A1}$$

but

$$C_{TFWT} = C_{air} * \frac{RF_p}{H_a} = C_{am} * RF_p \tag{A2}$$

Consequently, the following equation for OBT concentration in plants is obtained:

$$C_{OBT} = DW_p * ID_p * WE_p * C_{TFWT} \tag{A3}$$

where: $RF_p$ is the reduction factor used because tritium concentration in soil is much smaller than that in air (unit less); $DW_p$ is the dry matter (dm) fraction of plant (kg dry weight (dw) plant kg$^{-1}$ fresh weight (fw) plant); $ID_p$ is the isotopic discrimination factor due to plant physiology (unit less); $WE_p$ is the water equivalent of organic matter (at combustion in OBT measurements); $H_a$ is the absolute atmospheric humidity (L m$^{-3}$); $C_{air}$ is the HTO concentration in air (Bq m$^{-3}$); $C_{am}$ is the HTO concentration in air moisture (Bq m$^{-3}$); $C_{TFWT}$ is the HTO concentration in the leaf free water (Bq L$^{-1}$)

The IAEA approach (IAEA, 2009) uses an equation similar to equation (A3) for OBT concentration in plants, but with somewhat different notation:

$$C_{OBT} = (1 - WC_p) * WEQ_p * R_p * C_{TFWT} \tag{A4}$$

where: $WEQ_p$ is the water equivalent factor (kg of water produced per kg dm combusted) - $WEQ_p$ in equation (A4) is the same as $WE_p$ in equation (A3); $WC_p$ is the fractional water content of the plant (L kg$^{-1}$) – $1- WC_p$ in equation (A4) is the same as $DW_p$ in equation (A3); $R_p$ is the partition factor for plants – $R_p$ in equation (A4) is the same as $ID_p$ in equation (A3).

IAEA (2009) recommends the following equation for tritium concentration in plant water, with an explicit distinction between air and root soil pathway:

$$C_{TFWT} = \frac{RH*C_{am}+(1-RH)*C_{sw}}{\gamma} \quad (A5)$$

where: $C_{TFWT}$ is the HTO concentration in the leaf free water (Bq L$^{-1}$); $RH$ is the relative humidity of air; $C_{am}$ is the HTO concentration in air moisture (Bq L$^{-1}$); $C_{sw}$ is the HTO concentration in the soil water (Bq L$^{-1}$); $\gamma$ (= 0.909) is the ratio of the HTO vapour pressure to that of H$_2$O.

The ratio between soil water, $C_{sw}$ and air moisture, $C_{am}$ can vary between various sites and a default value of 0.3 is recommended by IAEA (2009, 2010).

For equilibrium conditions and considering OBT as water of combustion, the following equations are obtained:

CSA (2014): $\quad \frac{OBT}{TFWT} = ID_p \quad$ (total OBT) $\quad$ (A6)

IAEA (2009): $\quad \frac{OBT}{TFWT} = R_p \quad$ (NE-OBT) $\quad$ (A7)

The values of the isotopic discrimination factor, $ID_p$ given in various studies (McFarlane, 1976; Garland and Ameen, 1979; Kim and Baumgartner, 1988; Hisamatsu et al., 1989; Inoue and Iwakura, 1990; Kotzer and Workman, 1999) range from 0.64 to 1.3. Based on that narrow range, the arithmetic mean of 0.8 should be used as the default value (CSA, 2014a).The partition factor, $R_p$ accounts for the reduction in dry weight (dw) concentration due to the presence of exchangeable hydrogen in combustion water, as well as for isotopic discrimination factor. The values of the partition factor, $R_p$ must be determined empirically for steady-state conditions. Although the number of data points is small, all values are less than one, with a GM of 0.54 for the crops considered (maize, barley and alfalfa). In the absence of other information, this value is assumed to apply to all plant types. Regardless of the plant in question, the TFWT concentration used in Equation (A4) should be the concentration in the plant leaves, the primary location of dry matter production (IAEA, 2009).

**References**

Amano, H., 1995. Preliminary measurement on uptake of tritiated methane by plants. Fusion Technol. 28, 797-802.

ASN, 2010. Livre blanc du tritium. Autorite de Surete Nucleaire (in French). http://www.asn.fr/sites/tritium/fichiers/Tritium_livre_blanc_integral_web.pdf.

Barry, P.J., Watkins, B.M., Belot, Y., Etlund, O., Galeriu, D., Raskob, W., Russel, S., Togawa, O., 1999. Intercomparison of model predictions of tritium concentration in soil and foods following acute airborne HTO exposure. J. Environ. Radioact. 42, 191-207.

Belot, Y., Camus, H., Marini, T., 1992. Determination of tritiated formaldehyde in effluents from tritium facilities. Fusion Technol. 21, 556-559.

Belot, Y., Camus, H., Marini, T., Raviart, S., 1993. Volatile tritiated organic acids in stack effluents and in air surrounding contaminated materials. J. Fusion. Energ. 12, 71-75.

Belot, Y., Camus, H., Raviart, S., Antoniazzi, A.B., Shmayda, W.T., 1995. Production of tritiated organic acids at tritium-bearing stainless steel surfaces exposed to air. Fusion Technol. 28, 1138-1143.

Brown, J., Simmonds, J.R., 1995. FARMLAND – A dynamic model for the transfer of radionuclides through terrestrial foodchains. NRPB-R-273. National Radiological Protection Board, Chilton, UK.

Caird, M.A., Richards, J.H., Donovan, L.A., 2007. Nighttime stomatal conductance and transpiration in C3 and C3 plants. Plant Physiol. 143, 4-10.

CERC, 2008. ADMS Version 4. Cambridge Environmental Research Consultants. http://www.cerc.co.uk/

Clark, I., Mihok, S., Wilk, M., Lapp, A., Dehay-Turner, B., St-Amant, N., Kwamena, N., 2014. HTO – OBT dynamics in plants grown in a high HT environment. 3$^{rd}$ Organically Bound Tritium (OBT) Workshop, September 15 -18, 2014, Ottawa, Canada.

CNSC, 2013. Environmental fate of tritium in soil and vegetation. Part of the Tritium Studies Project. Canadian Nuclear Safety Commission. https://www.cnsc-ccsn.gc.ca/eng/pdfs/Reading-Room/healthstudies/Environmental-Fate-of-Tritium-in-Soil-and-Vegetation-eng.pdf.

CSA, 1987. Guidelines for calculating derived release limits for radioactive material in airborne and liquid effluents for normal operation of nuclear facilities. Canadian Standards Association Group. CAN/CSA N288.1-M87, Mississauga, Canada.

CSA, 2008. Guidelines for calculating derived release limits for radioactive material in airborne and liquid effluents for normal operation of nuclear facilities. Canadian Standards Association Group. CSA N288.1-08, Mississauga, Canada.

CSA, 2014a. Guidelines for calculating derived release limits for radioactive material in airborne and liquid effluents for normal operation of nuclear facilities. Canadian Standards Association Group. CSA N288.1-14, Mississauga, Canada.

CSA, 2014b. Guidelines for calculating the radiological consequences to the public of a release of airborne radioactive material for nuclear reactor accidents. Canadian Standards Association Group. CSA N288.2-14, Mississauga, Canada.

Davis, P.A., Kim, S.B, Chouhan, S.L., Workman, W., J., G., 2005. Observed and modeled tritium concentrations in the terrestrial food chain near a continuous atmospheric source.Fusion Sci. Technol. 48, 504-507.

Diabate, S., Strack, S., 1993. Organically bound tritium. Health Phys. 65, 698-712.

Dunne, T., Malmon, D.V., Mudd, S.M., 2010. A rain splash transport equation assimilating field and laboratory measurements. J. Geophys. Res. 115, F01001, doi:10.1029/2009JF001302

EA, 2002. Principles for the assessment of prospective public doses arising from authorised discharges of radioactive waste to the environment. Radioactive Substances Regulation under the Radioactive Substances Act (RSA-93) or under the Environmental Permitting Regulations (EPR-10). Environment Agency, UK. https://www.gov.uk/government/uploads/system/uploads/attachment_data/file/296390/geho1202bklh-e-e.pdf. http://publications.environment-agency.gov.uk/pdf/PMHO1202BKLH-e-e.pdf.

Fairlie, I., 2010. Hypothesis to explain childhood cancer near nuclear power plants. Int. J. Occup. Environ. Health 16, 341–350.

Galeriu, D., 1994. Transfer parameters for routine release of HTO − Consideration of OBT. AECL-11052. Atomic Energy of Canada Limited, Chalk River, Ontario, Canada.

Galeriu, D., Raskob, W., Melintescu, A., Turcanu, C., 2000a. Model description of the Tritium Food Chain and Dose Module FDMH in RODOS PV 4. RODOS (WG3)-TN(99)-54.

Galeriu, D., Raskob, W., Melintescu, A., Turcanu, C., 2000b. Documentation of tritium food chain and dose module FDMH in RODOS PV4. RODOS(WG3)-TN(99)-56.

Galeriu, D., Melintescu, A., Slavnicu, D., 2009. IFIN - Cernavoda NPP: Consultancy on Canadian Standard CSA-N-288.1-2008 in conformity with norms in European Community and the IAEA guidance for assessing Derived Release Limits for $^3$H and $^{14}$C (in Romanian)

Galeriu, D., Melintescu, A., Slavnicu, D., Gheorghiu, D., Simionov, V., 2009. Accidental release of tritiated water - toward a better radiological assessment. Radioprotection 44, 177 – 183.

Galeriu, D., Melintescu, A., 2010. Tritium. In: Atwood, D.A. (Ed.), Radionuclides in the environment. John Wiley & Sons Ltd., West Sussex, England, pp. 47-65.

Galeriu, D., Melintescu, A., 2012. Research and development of environmental tritium modelling, an update. The 57[th] Annual Meeting of the Health Physics Society, 22-26 July 2012, Sacramento, CA.
http://hpschapters.org/sections/envrad/2012AMpresentations/HP2012Galeriu.pdf.

Galeriu, D., Melintescu, A., Strack, S., Atarashi-Andoh, M., Kim, S.B., 2013. An overview of organically bound tritium experiments in plants following a short atmospheric HTO exposure. J. Environ. Radioact. 118, 40-56.

Galeriu, D., Melintescu, A., 2016. Relevance of night production of OBT in crops. Oral presentation at 11[th] International Conference on Tritium Science and Technology (TRITIUM 2016) (http://www.ans.org/meetings/m_213, http://tritium2016.org/), accepted to Fusion Sci. Technol. FST16-174R1.

Garland, J.A., Ameen, M., 1979. Incorporation of tritium in grain plants. Health Phys. 36, 35-38.

Goudriaan, J., van Laar, H.H., 1994. Modelling potential crop growth processes. Kluwer Academic Publishers, Dordrecht, The Netherlands.

Guetat, P., Patryl, L., 2008. Environmental and radiological impact of accidental tritium release. Fusion Sci. Technol. 54, 273-276.

Higgins, N.A., 1997. TRIF — An intermediate approach to environmental tritium modelling. J. Environ. Radioact. 36, 253-267.

Hisamatsu, S., Takizawa, Y., Itoh, M., Ueno, K., Katsumata, T., Sakanoue, M., 1989. Fallout $^3$H in human tissue at Akita, Japan. Health Phys. 57, 559-563.

IAEA, 2003. Modelling the environmental transport of tritium in the vicinity of long term atmospheric and sub-surface sources. IAEA-BIOMASS-3. International Atomic Energy Agency, Vienna.

IAEA, 2008a. Modelling the Environmental Transfer of Tritium and Carbon-14 to Biota and Man. Report of the Tritium and Carbon-14 Working Group of EMRAS Theme 1. Environmental Modelling for RAdiation Safety (EMRAS) Programme. International Atomic Energy Agency Vienna. http://www-ns.iaea.org/downloads/rw/projects/emras/draft-final-reports/emras-tritium-wg.pdf.

IAEA, 2008b. The Pickering Scenario. Final report. Working Group on Modelling of Tritium and Carbon-14 Transfer to Biota and Man. Environmental Modelling for RAdiation Safety (EMRAS) Programme. International Atomic Energy Agency Vienna. http://www-ns.iaea.org/downloads/rw/projects/emras/tritium/pickering-final.pdf.

IAEA, 2009. Quantification of radionuclide transfer in terrestrial and freshwater environments for radiological assessments. IAEA-TECDOC-1616. International Atomic Energy Agency, Vienna.

IAEA, 2010. Handbook of parameter values for the prediction of radionuclide transfer in terrestrial and freshwater environments. Technical reports series no. 472. IAEA-TRS-472. International Atomic Energy Agency, Vienna.

IAEA, 2014a. Transfer of tritium in the environment after accidental releases from nuclear facilities. Report of Working Group 7 of the IAEA's Environmental Modelling for Radiation Safety (EMRAS II) Programme. IAEA-TECDOC-1738. International Atomic Energy Agency Vienna

IAEA, 2014b. Radiation protection and safety of radiation sources: International basic safety standards. General safety requirements Part 3 No. GSR Part 3. International Atomic Energy Agency Vienna.

ICRP, 2006. Assessing dose of the representative person for the purpose of the radiation protection of the public. ICRP Publication 101a, Annals of the ICRP 36 (3). International Commission on Radiological Protection, Oxford, Elsevier Science.

Ichimasa, M., Suzuki, M., Obayashi, H., Sakuma, Y., Ichimasa, Y., 1999. *In vitro* determination of oxisation of atmospheric tritium gas in vegetation and soil in Ibaraki and Gifu, JAPAN. J. Radiat. Res. 40, 243-251.

Imboden, S.F., Overcamp, T.J., 2006. Chronic dose due to a continuous tritium release calculated by CAP88-PC and NORMTRI. Nucl. Technol. 155, 114-118.

Inoue, Y., Iwakura, T., 1990. Tritium concentration in Japanese rice. J. Radiat. Res. 31, 311-323.

Kakiuchi, H., Andoh, M.A., Amano, H., 2002. The evaluation of uptake of tritiated methane to plants. International symposium: Transfer of radionuclides in biosphere. Prediction and assessment, Mito, Ibaraki, Japan, 18 – 19 December 2002. http://www.iaea.org/inis/collection/NCLCollectionStore/_Public/36/116/36116269.pdf.

KEPCO, 2012. Environmental monitoring and evaluation report around nuclear power plants. Annual report. Korea Electric Power Corporation (in Korean).

KEPCO, 2014. Environmental monitoring and evaluation report around nuclear power plants. Annual report. Korea Electric Power Corporation (in Korean).

Kim, M.-A., Baumgartner, F., 1988. Validation of tritium measurements in biological materials. Fusion Technol. 14, 1153-1156.

Kim, S.B., Baglan, N., Davis, P.A., 2013. Current understanding of organically bound tritium (OBT) in the environment. J. Environ. Radioact. 126, 83-91.

Kim, S.B., Workman, W.J.G., Davis, P.A., 2008. Experimental investigation of buried tritium in plant and animal tissues. Fusion Sci. Technol. 54, 257-260.

Kim, S.B., Korolevych, V., 2013. Quantification of exchangeable and non-exchangeable organically bound tritium (OBT) in vegetation. J. Environ. Radioact. 118, 9-14.

Korolevych, V.Y, Kim, S.B., 2013. Relation between the tritium in continuous atmospheric release and the tritium contents of fruits and tubers. J. Environ. Radioact. 118, 113-120.

Korolevych, V.Y., Kim, S.B., Davis, P.A., 2014. OBT/HTO ratio in agricultural produce subjected to routine atmospheric releases of tritium. J. Environ. Radioact. 129, 157-168.

Kotzer, T.C., Workman, W.J.G., 1999. Measurements of tritium (HTO, TFWT, OBT) in environmental samples at varying distances from a Nuclear Generation Station. Report No. AECL-12029. Chalk River, Atomic Energy of Canada Limited

Maro, D., Vermorel, F., Rozet, M., Aulagnier, C., Hebert, D., Le Dizes, S., Voiseux, C., Solier, L., Cossonet, C., Godinot, C., Fievet, B., Laguionie, P., Connan, O., Cazimajou, O., Morillo, M, Lamothe, M., 2016. The VATO project: an original methodology to study the transfer of tritium as HT and HTO in grassland ecosystem. Submitted to J. Environ. Radioact.

Mayall, A., Cabianca, T., Attwood, C.A., Fayers, C.A., Smith, J.G., Penfold, J., Steadman, D., Martin, G., Morris, T.P., Simmonds, J.R., 1997. PC-CREAM installing and using the PC system for assessing the radiological impact of routine releases. EUR 17791 EN, NRPB-SR296, Chilton, UK.

McFarlane, J.C., 1976. Tritium fractionation in plants. Environ. Exp. Bot. 16, 9-14.

Melintescu, A., Galeriu, D., Marica, E., 2002. Using WOFOST crop model for data base derivation of tritium and terrestrial food chain modules in RODOS. Radioprotection 37, 1242-1246.

Melintescu, A., Galeriu, D., 2005. A versatile model for tritium transfer from atmosphere to plant and soil. Radioprotection 40, S437-S442.

Melintescu, A., Galeriu, D., 2011. Exchange velocity approach and the role of photosynthesis for tritium transfer from atmosphere to plants. Fusion Sci. Technol. 60, 1179 – 1182.

Melintescu, A., Galeriu, D., Diabaté, S., Strack, S., 2015. Preparatory steps for a robust dynamic model for OBT dynamics in agricultural crops. Fusion Sci. Technol. 67, 479-482.

Michelotti, E., Green, A., Whicker, J., Eisele, W., Fuehne, D., McNaughton, M., 2013. Validation test for CAP88 predictions of tritium dispersion at Los Alamos National Laboratory. Health Phys., 105, S176-S181.

Mihok, S., Wilk, M., Lapp, A., St-Amant, N., Kwamena, N.-O.A., Clark, I.D., 2016. Tritium dynamics in soil and plants grown under three irrigation regimes at a tritium processing facility in Canada. J. Environ. Radioact. 153, 176-187.

Mitchell, N.G., 1999. A handbook for the SPADE suite of models for radionuclides transfer through terrestrial foodchains. In: Mouchel Report 48112 produced for Ministry of Agriculture, Fisheries and Food/Food Standard Agency, UK.

Murphy Jr., C.E., 1993. Tritium transport and cycling in the environment. Health Phys. 65, 683-697.

NDAWG, 2011. Short-term releases to the atmosphere. NDAWG/2/2011. National Dose Assessment Working Group, UK.

Ota, M., Yamazawa, H., Moriizumi, J., Iida, T., 2007. Measurement and modeling of oxidation rate of hydrogen isotopic gases by soil. J. Environ. Radioact. 97, 103-115.

Ota, M., Nagai, H., 2011. Development and validation of a dynamical atmosphere – vegetation – soil HTO transport and OBT formation model. J. Environ. Radioact. 102, 813-823.

Ota, M., Nagai, H., 2012. HTO transport and OBT formation after nighttime wet deposition of atmospheric HTO onto land surface. Radioprotection, 46, S417-S422.

Patryl, L., Armand, P. 2005. Modelisation des tranferts du tritium a l'environnement. Technical report, E.O.T.P. A-2S000-01-21-31-01. Commissariat a l'Energie Atomique, Bruyeres-le-Chatel, France (in French).

Patryl, L., Armand, P., 2007. Key assumptions and modelling approaches of tritium models implemented in GAZAXI 2002 and CERES. CEA report EOTP A-24100-01-01-AW-43.

Paunescu, N., Galeriu, D., Mocanu, N., 2002. Environmental tritium around a new CANDU nuclear power plant. Radioprotection, 37, C1-1253-C1-1258.

Peterson, S.R., Hoffman, F.O., Köhler, H., 1996. Summary of the BIOMOVS A4 scenario: Testing models of the air-pasture-cow milk pathway using Chernobyl fallout data. Health Phys. 71, 149-159.

Raskob, W., 1993. Description of the new version 4.0 of the tritium model UFOTRI including user guide. Report KfK 5194. Kernforschungszentrum Karlsruhe, Germany.

Raskob, W., 1994. NORMTRI, A computer program for assessing the off-Site consequences from air-borne releases of tritium during normal operation of nuclear facilities. Report KfK-5364. Kernforschungszentrum Karlsruhe, Germany.

Raskob, W., Diabate, S., Strack, S., 1996. A new approach for modelling the formation and translocation of organically bound tritium in accident consequence assessment codes. Int. Symposium on Ionization Radiation: Protection of the Natural Environment, Stockholm, Sweden, May 20–24, 1996.

Raskob, W., 2011. Plant OBT model. The Fifth Meeting (TM) on the Environmental Modelling for Radiation Safety (EMRAS II) Intercomparison and Harmonization Project, January 24 – 28, 2011, Vienna, Austria. http://www-ns.iaea.org/downloads/rw/projects/emras/emras-two/third-technical-meeting/wgroup-seven/presentation-5th-wg7-plant-obt-model.pdf.

Raskob, W., 2007. Test and validation studies performed with UFOTRI and NORMTRI. Report FZK 7281, TW5-TSS/SEP2 – deliverable 4. Forschungszentrum Karlsruhe, Germany.

Schulman, L.L., Strimaitis, D.G., Scire, J.S., 1997. The PRIME plume rise and building downwash model. Addendum to ICS3 user's guide. Earth Tech., Inc., Concord, MA. https://www3.epa.gov/ttn/scram/7thconf/iscprime/useguide.pdf.

Shen, H.-F., Liu, W., 2016. Tritium concentration in soybean plants exposed to atmospheric HTO during nighttime. Nucl. Sci. Tech. 27, 39, doi:10.1007/s41365-016-0031-8.

Simon-Cornu, M., Beaugelin-Seiller, K., Boyer, P., Calmon, P., Garcia-Sanchez, L., Mourlon, C., Nicoulaud, V., Sy, M., Gonze, M.A., 2015. Evaluating variability and uncertainty in radiological impact assessment using SYMBIOSE. J. Environ. Radioact. 139, 91-102.

St-Aman, N., 2016. An overview of tritium studies at the Canadian Nuclear Safety Commission. 5th Workshop on OBT (Organically Bound Tritium) and its analysis. Le Mans, France, 4 – 7 October 2016

Strack, S., Diabate, S., Raskob, W., 1998. Modellrechnungen zur Biokinetik von Tritium in Pflanzen, in: Winter, M. (Hrsg.), Radioaktivitat in Mensch und Umwelt, 30. FS Jahrestagung, 28.9. – 2.10.98, Lindau, Bd. II, S. 855-860, Koln: TueV Rheinland (in German).

Strack, S., Diabate, S., Raskob, W., 2005. Organically bound tritium in plants: insights gained by long-term experience in experimental and modelling research. Fusion Sci. Technol. 48, 767-770.

Strack, S., Diabate, S., Raskob, W., 2011. Biokinetics of tritium in wheat plants: measurements and model calculations. Sixth Meeting of the EMRAS II Working Group 7, "Tritium" Accidents, Bucharest, Romania, 12 – 15 September 2011. http://www-ns.iaea.org/downloads/rw/projects/emras/emras-two/first-technical-meeting/sixth-working-group-meeting/working-group-presentations/workgroup-7-presentations/presentation-6th-wg7-experimental-work.pdf.

Thompson, P.A., Kwamena, N.-O.A., Ilin, M., Wilk, M., Clark, I.D., 2015. Levels of tritium in soils and vegetation near Canadian nuclear facilities releasing tritium to the atmosphere: implications for environmental models. J. Environ. Radioact. 140, 105-113.

**FIGURE CAPTION**

**Fig. 1**  Simulation of HTO concentration in air during a year

**Fig. 2**  Simulation of HTO concentration in air during 24 hours period

**Fig.3**  OBT/HTO ratio in barley and rice (based on experimental data from KEPKO, 2012, 2014)

**Fig. 4**  OBT/HTO ratio in cabbage and persimmon (based on experimental data from KEPKO, 2012, 2014)

**Fig. 5**  OBT/HTO ratio in milk (based on experimental data from KEPKO, 2012, 2014)

**Fig. 6**  Ingestion dose of adult and infant for a spike release of one hour at mid-day during a year

**Fig. 7**  Experimental OBT/HTO ratio for wheat leaves at various exposures carried out at different hours of the day and night (s7, s8, s10, s11, s20p, s23p)

**Fig. 8**  Predicted to observed ratios (P/O) for concentrations of OBT in grain at harvest

**Fig. 9**  Predicted to observed ratios (P/O) for concentrations of OBT in leaves at the end of exposure

**Fig. 10**  Predicted to observed ratios (P/O) for concentrations of OBT in leaf at harvest

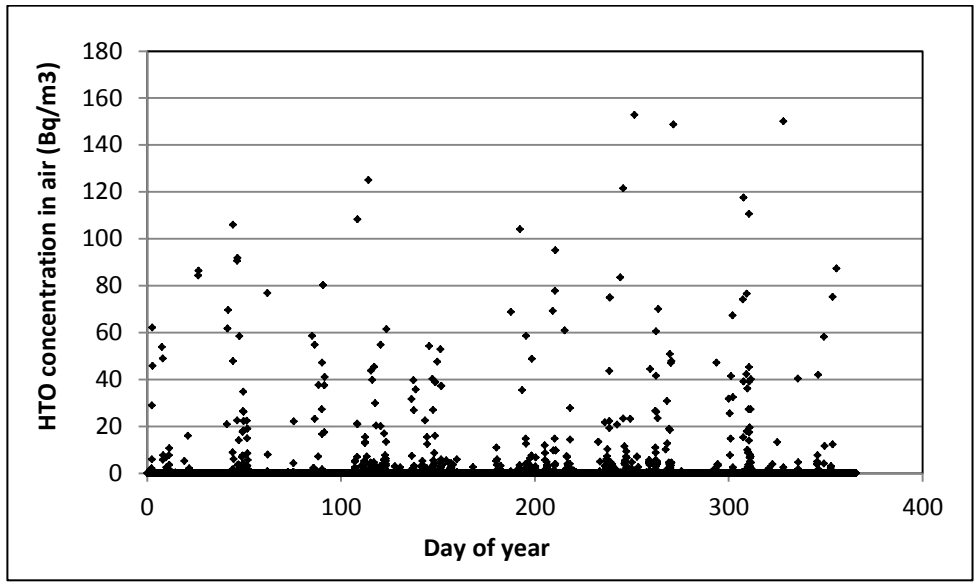

**Fig. 1**

**Fig. 2**

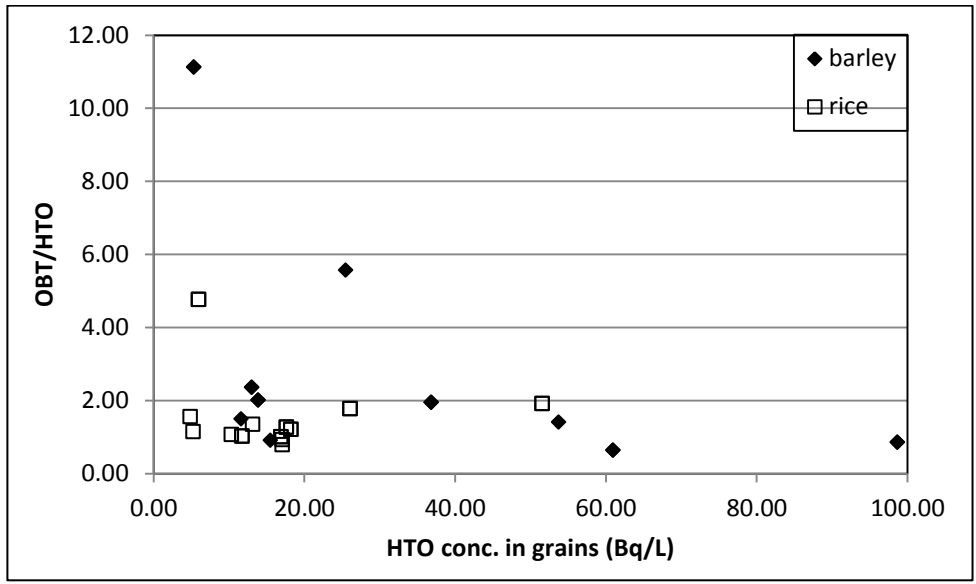

**Fig. 3**

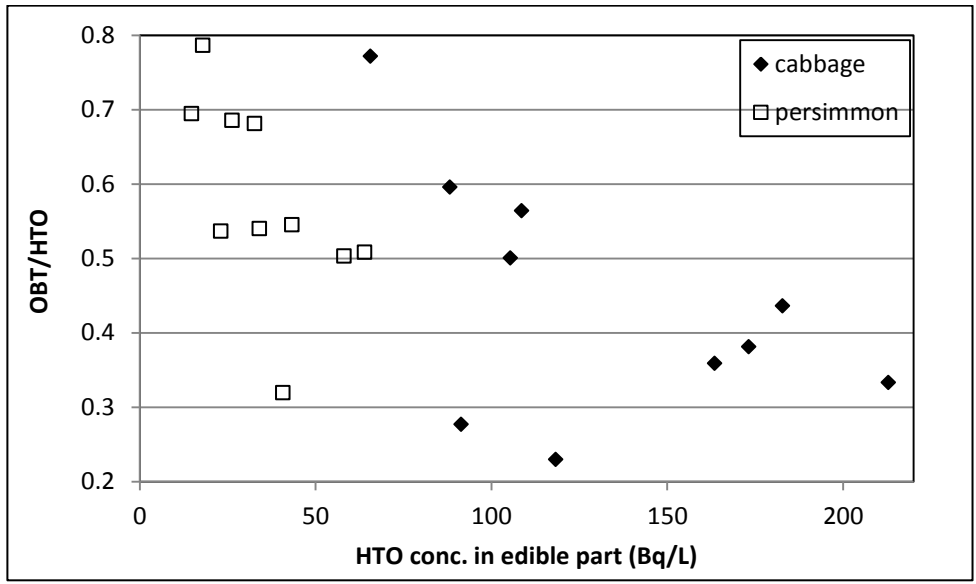

**Fig. 4**

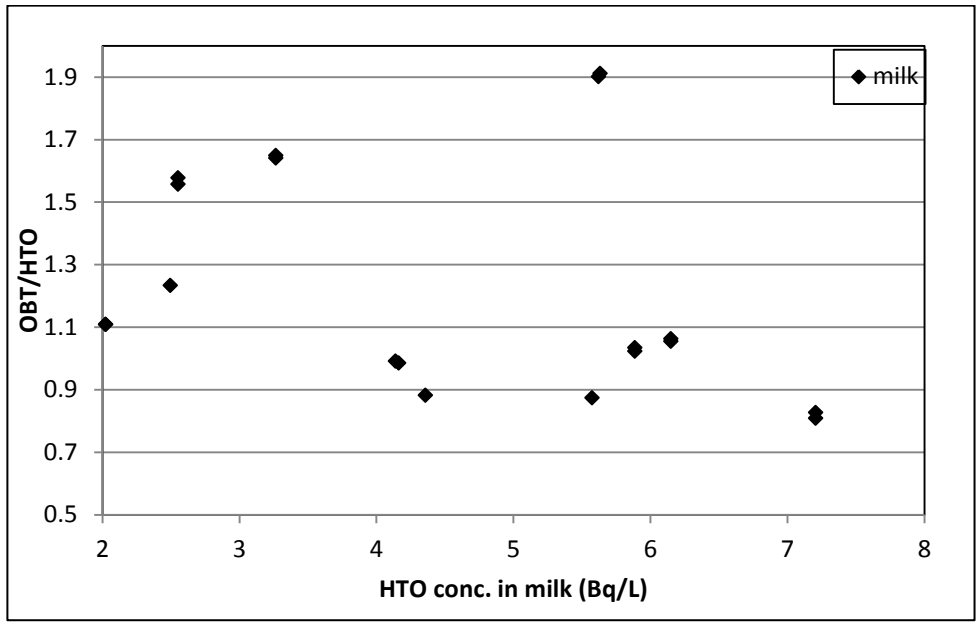

**Fig. 5**

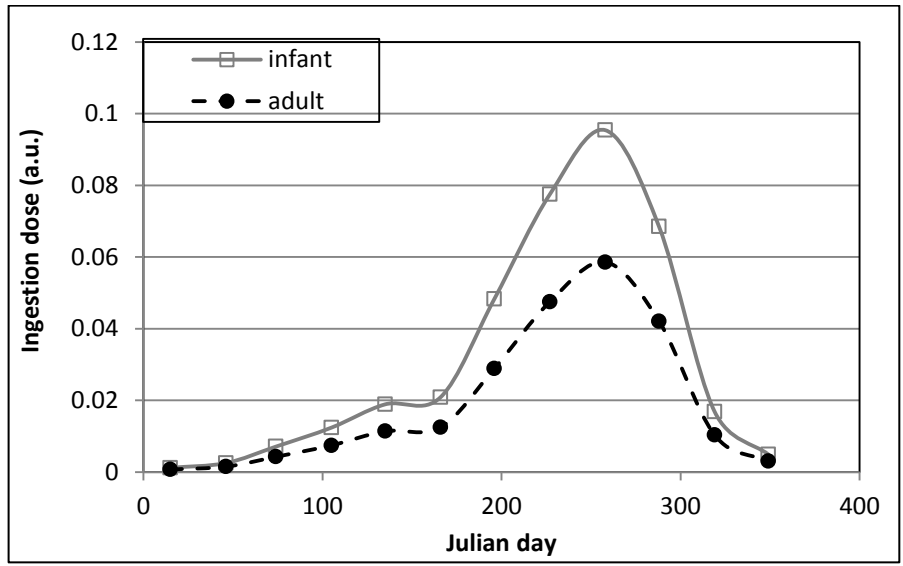

**Fig. 6**

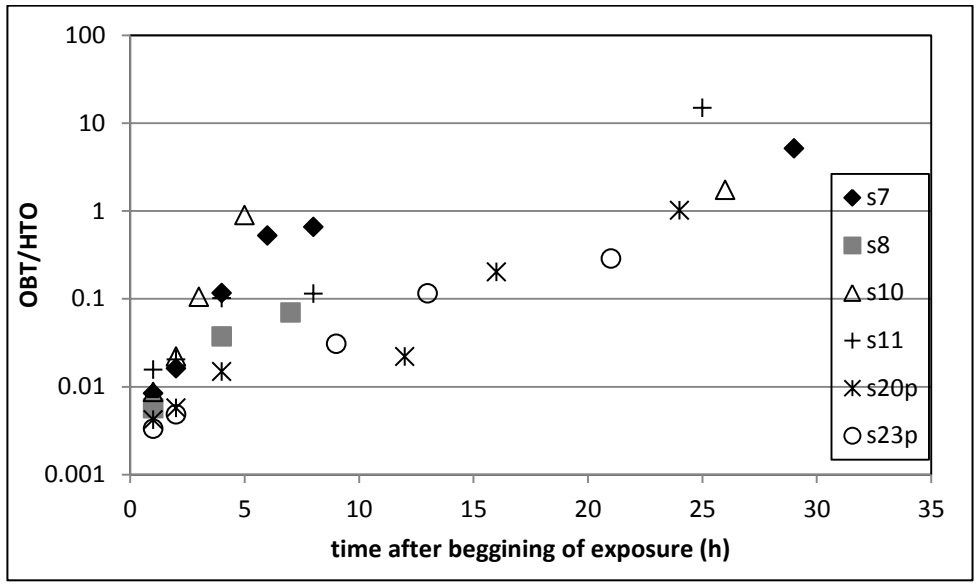

**Fig. 7**

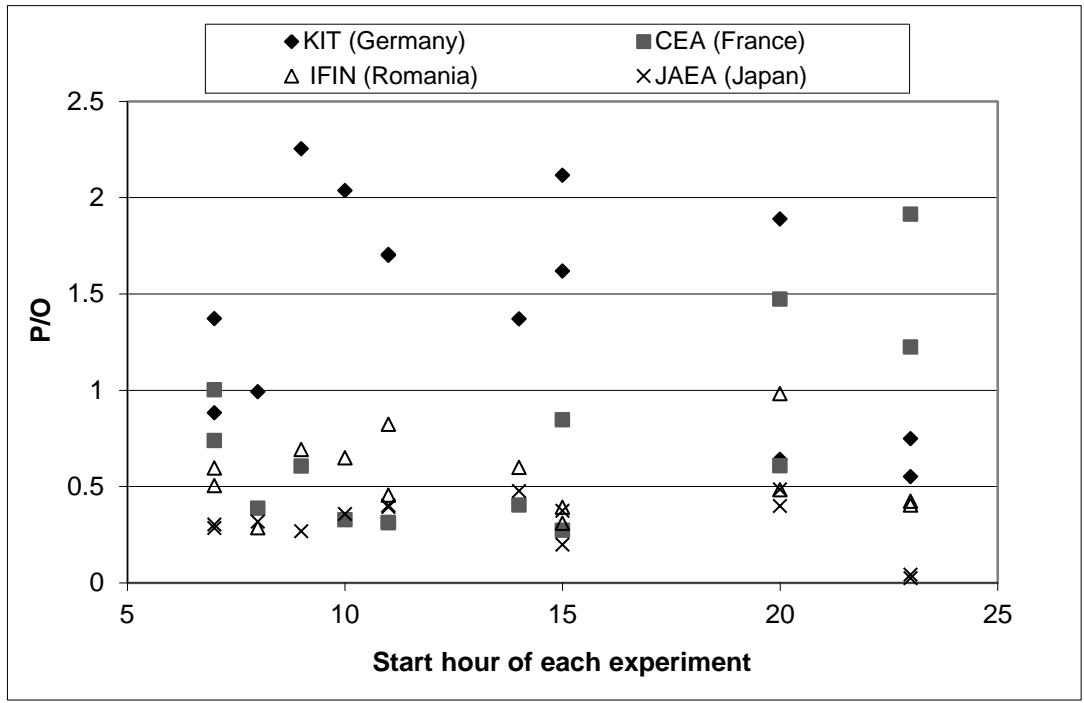

**Fig. 8**

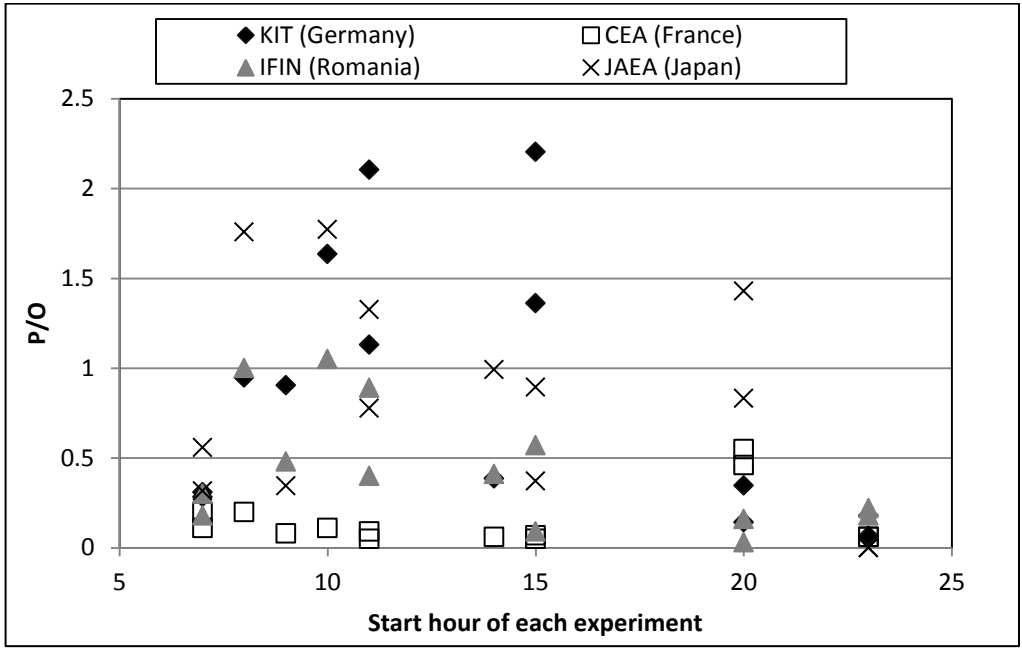

**Fig. 9**

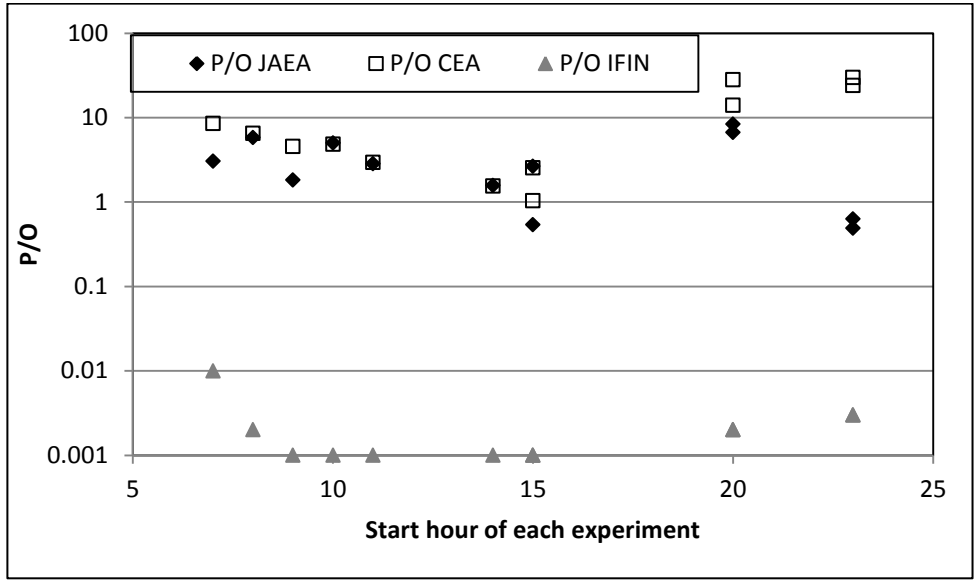

**Fig. 10**

**Table 1**

The ingestion doses (µSv year$^{-1}$) for infant, child and adult for an atmospheric tritium emission of 1 TBq for various atmospheric stability conditions and durations

| *Case* | *Normal day* | *Category D* | *Category F* | *Continuous release, 1 year* | *Category F/continuous release ratio* |
|---|---|---|---|---|---|
| Infant | 1.33 | 2.22 | 8.24 | 0.23 | 35.21 |
| Child | 1.68 | 2.77 | 10.20 | 0.24 | 41.80 |
| Adult | 1.80 | 3.96 | 14.30 | 0.26 | 55.21 |

**Table 2**

The ratio between tritium concentration in soil water and air moisture ($C_{sw}/C_{am}$) and the ratio between tissue free water tritium concentration in plants (TFWT) and air moisture ($C_{TFWT}/C_{am}$) for various values of relative humidity (RH)

| *RH* | 0.6 | 0.7 | 0.8 |
|---|---|---|---|
| $C_{sw}/C_{am}$ | | $C_{TFWT}/C_{am}$ | |
| 0.1 | **0.70** | **0.80** | **0.90** |
| 0.2 | **0.75** | **0.84** | **0.92** |
| 0.3 | **0.79** | **0.87** | **0.95** |
| 0.4 | **0.84** | **0.90** | **0.97** |
| 0.5 | **0.88** | **0.94** | **0.99** |
| 0.6 | **0.92** | **0.97** | **1.01** |

**Table 3**

Dry matter fraction (DW$_p$) of terrestrial plants

| Plants | GM[a] | GSD[b] | CV[c] | minimum | maximum |
|---|---|---|---|---|---|
| All leafy vegetables | 0.069 | 1.312 | 0.277 | 0.030 | 0.160 |
| Root vegetables | 0.107 | 1.281 | 0.252 | 0.050 | 0.230 |
| Fruits | 0.104 | 1.363 | 0.318 | 0.040 | 0.270 |
| Pasture | 0.182 | 1.214 | 0.196 | 0.100 | 0.330 |
| Cereals | 0.869 | 1.011 | 0.011 | 0.840 | 0.900 |
| Silage | 0.285 | 1.160 | 0.150 | 0.180 | 0.450 |

[a] geometric mean, [b] geometric standard deviation, [c] coefficient of variance

**Table 4**

Water equivalent factor (WE$_p$) of terrestrial plants

| Plants | GM$^a$ | GSD$^b$ | CV$^c$ | minimum | maximum |
|---|---|---|---|---|---|
| All leafy vegetables | 0.508 | 1.035 | 0.034 | 0.470 | 0.550 |
| All non-leafy vegetables | 0.524 | 1.021 | 0.021 | 0.500 | 0.550 |
| Root vegetables | 0.497 | 1.045 | 0.044 | 0.450 | 0.550 |
| Other vegetables | 0.548 | 1.040 | 0.040 | 0.500 | 0.600 |

$^a$ geometric mean, $^b$ geometric standard deviation, $^c$ coefficient of variance

**Table 5**

Public dose (Sv year$^{-1}$) coming from different tritium contamination pathways for children of 1 year and 10 years old and for an adult when OBT/HTO ratio is 0.7 and tritium concentration in animal drinking water is 4.5 Bq L$^{-1}$

| Contamination pathway | 1 y | 10 y | Adult |
|---|---|---|---|
| Inhalation | 1.97x10$^{-7}$ | 2.71x10$^{-7}$ | 2.27x10$^{-7}$ |
| Drinking water | 8.67x10$^{-8}$ | 5.84x10$^{-8}$ | 7.68x10$^{-8}$ |
| Food HTO | 6.17x10$^{-7}$ | 4.29x10$^{-7}$ | 3.32x10$^{-7}$ |
| Food OBT | 9.08x10$^{-8}$ | 9.79x10$^{-8}$ | 8.71x10$^{-8}$ |
| Total | 9.92 x10$^{-7}$ | 8.57x10$^{-7}$ | 7.23x10$^{-7}$ |

**Table 6**

Public dose (Sv year$^{-1}$) coming from different tritium contamination pathways for children of 1 year and 10 years old and for an adult when OBT/HTO ratio is 0.7 and tritium concentration in animal drinking water is 40 Bq L$^{-1}$

| Contamination pathway | 1 y | 10 y | Adult |
|---|---|---|---|
| Inhalation | 1.97x10$^{-7}$ | 2.71x10$^{-7}$ | 2.27x10$^{-7}$ |
| Drinking water | 8.67x10$^{-8}$ | 5.84x10$^{-8}$ | 7.68x10$^{-8}$ |
| Food HTO | 1.08x10$^{-6}$ | 5.76x10$^{-7}$ | 4.48x10$^{-7}$ |
| Food OBT | 1.39x10$^{-7}$ | 1.15x10$^{-7}$ | 9.87x10$^{-8}$ |
| Total | 1.50x10$^{-6}$ | 1.02x10$^{-6}$ | 8.50x10$^{-7}$ |

**Table 7**

Public dose (Sv year$^{-1}$) coming from different tritium contamination pathways for children of 1 year and 10 years old and for an adult when OBT/HTO ratio is 10 and tritium concentration in animal drinking water is 4.5 Bq L$^{-1}$

| Contamination pathway | 1 y | 10 y | Adult |
|---|---|---|---|
| Inhalation | 1.97x10$^{-7}$ | 2.71x10$^{-7}$ | 2.27x10$^{-7}$ |
| Drinking water | 8.67x10$^{-8}$ | 5.84x10$^{-8}$ | 7.68x10$^{-8}$ |
| Food HTO | 6.17x10$^{-7}$ | 4.29x10$^{-7}$ | 3.32x10$^{-7}$ |
| Food OBT | 1.05x10$^{-6}$ | 1.19x10$^{-6}$ | 1.07x10$^{-6}$ |
| Total | 1.95x10$^{-6}$ | 1.95x10$^{-6}$ | 1.70x10$^{-6}$ |

**Table 8**

Comparison between NDAWG results (2011) and TRIF model results (Higgins, 1997) for integrated activity (Bq d kg $^{-1}$) of various food items in case of a normal day (class D, 12 hours)

| Crop | NDAWG | TRIF |
|---|---|---|
| Green vegetables | $7.29 \times 10^3$ | $4.52 \times 10^2$ |
| Fruit | $7.29 \times 10^3$ | NA[a] |
| Root vegetables | $7.29 \times 10^3$ | $4.52 \times 10^2$ |
| Cow milk | $4.08 \times 10^2$ | $2.07 \times 10^2$ |
| Cow meat | $3.52 \times 10^2$ | $1.79 \times 10^2$ |
| Cow liver | $3.52 \times 10^2$ | NA[a] |
| Sheep meat | $5.36 \times 10^2$ | $2.73 \times 10^2$ |
| Sheep liver | $5.36 \times 10^2$ | NA[a] |

[a] not available

**Table 9**

Comparison of model predictions for ingestion dose (µSv year$^{-1}$) of adults for an atmospheric tritium emission of 1 TBq for various atmospheric stability conditions and durations

| Model | Normal day | Category D | Category F | Continuous release, 1 year | Category F/continuous release ratio |
|---|---|---|---|---|---|
| UFOTRI[a] | 1.10 | 1.30 | 2.7 | 0.03 | 90.20 |
| IFIN-HH[b] | 0.40 | 0.80 | 1.00 | 0.11 | 9.10 |
| NDAWG | 1.80 | 3.96 | 14.30 | 0.26 | 55.21 |

[a] NORMTRI used for continuous release; [b] CSA used for continuous release